\documentclass[11pt,a4paper]{article}
\usepackage[T1]{fontenc}
\usepackage[utf8]{inputenc}
\usepackage{lmodern}
\usepackage[margin=2.6cm]{geometry}
\usepackage{amsmath,amssymb,amsthm,mathtools}
\usepackage{booktabs,array}
\usepackage{graphicx}
\usepackage{tikz}
\usetikzlibrary{arrows.meta}
\usepackage{pgfplots,pgfplotstable}
\pgfplotsset{compat=1.18}
\usepackage{caption}
\usepackage{float}
\usepackage{fancyhdr}
\usepackage[numbers,square,sort&compress]{natbib}
\usepackage[colorlinks=true,linkcolor=blue!45!black,citecolor=blue!45!black,urlcolor=blue!45!black]{hyperref}
\hypersetup{
  pdftitle={Ordering Information in Multispecies Long-Range Swap Processes},
  pdfauthor={Eunghyun Lee},
  pdfsubject={Ordering information, Poisson coordinates and algebraic decay},
  pdfkeywords={mutual information, multispecies TASEP, Poisson processes, ordering, asymptotics}}

\newtheorem{theorem}{Theorem}[section]
\newtheorem{proposition}[theorem]{Proposition}
\newtheorem{lemma}[theorem]{Lemma}
\newtheorem{corollary}[theorem]{Corollary}
\theoremstyle{definition}
\newtheorem{definition}[theorem]{Definition}
\theoremstyle{remark}
\newtheorem{remark}[theorem]{Remark}

\newcommand{\Z}{\mathbb{Z}}
\newcommand{\R}{\mathbb{R}}
\newcommand{\E}{\mathbb{E}}
\newcommand{\PP}{\mathbb{P}}
\newcommand{\X}{\mathcal{X}}
\newcommand{\one}{\mathbf{1}}
\newcommand{\Id}{\mathrm{Id}}
\newcommand{\Var}{\operatorname{Var}}
\newcommand{\Cov}{\operatorname{Cov}}
\newcommand{\tr}{\operatorname{tr}}
\newcommand{\Pois}{\operatorname{Pois}}
\newcommand{\rank}{\operatorname{rank}}
\newcommand{\hb}{h_{\mathrm b}}
\newcommand{\TV}{\mathrm{TV}}
\newcommand{\LR}{\mathrm{LR}}
\newcommand{\ord}{\mathrm{ord}}
\newcommand{\T}{\mathsf{T}}
\definecolor{lrblue}{RGB}{24,98,140}
\definecolor{ordred}{RGB}{176,70,36}

\title{\textbf{Ordering Information in Multispecies\\ Long-Range Swap Processes}}
\author{Eunghyun Lee\\[2pt]
{\small School of Computing and Artificial Intelligence}\\
{\small Nazarbayev University, Astana, Kazakhstan}\\
{\small Correspondence: \texttt{eunghyun.lee@nu.edu.kz}}}
\date{}

\begin{document}
\maketitle
\thispagestyle{fancy}

\begin{abstract}
\noindent
We measure how much the left-to-right species order $W_t$ of interacting particles
on the integer lattice reveals about their initial configuration. For distinct
species under long-range swaps, deleting stronger particles gives coordinates with
independent Poisson increments. For a finitely supported initial law, the mutual
information between any function $U$ of the initial configuration and $W_t$ is
$C_U/t+O(t^{-3/2})$ as $t\to\infty$, where $C_U$, which may vanish, is determined
by the means of Gaussian order statistics and the conditional means of the initial
compressed coordinates given $U$. On consecutive sites with a uniform initial
order, we compute the coefficient of the $t\log(1/t)$ term in the short-time
information loss for arbitrary species counts. For three particles in this
setting, repeating the weaker rather than the stronger species triples the
long-time $t^{-1}$ coefficient, although both compositions have the same
composition entropy and initial-order entropy. These long-time coefficients are
the same whether equal particles block or push each other, but the nonuniform
limiting output law is reversed. In the ordinary multispecies totally asymmetric
simple exclusion process (TASEP), two distinct particles at fixed sites with a
uniform initial order instead give $t^{-1/2}$ decay.
\end{abstract}


\section{Introduction}

The \emph{species word} of a finite particle configuration on $\Z$ lists
the species labels from left to right. It records their relative order
and omits the particle positions and the gaps between successive particles.
Let $W_t$ denote this word at time $t$. How much information does $W_t$
retain about $W_0$ as time passes? When species labels repeat, how does
the rate of information loss depend on the number of particles of each
species? We label species from $1$ (weakest) to $q$
(strongest). If $n_i$ denotes the number of particles of species $i$,
then the ordered vector $(n_1,\ldots,n_q)$ is the \emph{composition},
and $n=\sum_{i=1}^q n_i$ is the total number of particles.

We consider two versions of the long-range swap dynamics
\cite{Lee2026,LeeJSTAT2026}. Each particle has an independent rate-one
Poisson clock. When its clock rings, the particle scans the sites to its
right in order. The scan skips stronger particles. Under the TASEP-type
rule, an encounter with a particle of the same species stops the scan
and leaves the configuration unchanged. Under the drop--push-type rule,
the scan continues past particles of the same species. In either version,
if the scan reaches a vacancy, the active particle moves there. If it
reaches a weaker particle, the two particles exchange positions. With a
single species, these rules give the totally asymmetric simple exclusion
process (TASEP) and the drop--push process, respectively. When all
particles have distinct species labels, the two long-range swap rules
coincide.

An initial configuration specifies the occupied sites and the species
label at each one. For a fixed composition, we draw the initial
configuration from a given law, independently of the clocks. This law
and the chosen rule determine the joint distribution of $W_0$ and $W_t$.
Their mutual information \cite{CoverThomas} is
\[
 I(W_0;W_t)=H(W_0)-H(W_0\mid W_t),
\]
where $H(W_0)$ is the Shannon entropy of the initial word and
$H(W_0\mid W_t)$ is its conditional entropy given the word at time $t$.
Thus $I(W_0;W_t)$ is the average reduction in uncertainty about the
initial word upon observing $W_t$. At $t=0$ it equals $H(W_0)$.
For each $t\geq 0$, it is zero if and only if $W_0$ and $W_t$ are
independent. We use natural logarithms and measure information in nats.
More generally, we study $I(U;W_t)$ for any random variable $U$
determined by the initial configuration. In particular, $U$ may be
the initial word or the initial configuration itself.

\subsection{Main results}

Theorem~\ref{thm:main} gives the long-time information for distinct species and arbitrary
initial inputs. Theorem~\ref{thm:repeated} treats the first two compositions with a
repeated species, and Theorem~\ref{thm:composition} gives the short-time expansion for
every composition.

\paragraph{Long times for distinct species.} Suppose that every species occurs once, so $q=n$ and the two rules coincide. To each particle we assign
a \emph{compressed coordinate}, its position minus the number of stronger particles to its left.
The map from a configuration to its vector of compressed coordinates is a bijection onto $\Z^n$ (Theorem~\ref{thm:poisson}).
When a clock rings, the coordinate of the ringing particle increases by one and all other
coordinates are unchanged (Theorem~\ref{thm:poisson}). The compressed coordinates therefore have independent
Poisson increments. The resulting product transition kernel (Corollary~\ref{cor:kernel})
also follows from the Bethe-ansatz formula of \cite[Theorem~3.14]{Lee2026}. Expanding the likelihood ratio between Poisson laws with different initial coordinates
at large $t$, we prove that, for every initial law $\lambda$
with finite support and every function $U$ of the initial configuration,
\[
 I(U;W_t)=\frac{C_U}{t}+O_{n,\lambda}(t^{-3/2})\qquad(t\to\infty).
\]
The constant is $C_U=\frac12\tr(J\Sigma_U)\ge0$. Here $\Sigma_U$ is the covariance matrix of
$m_U$, the conditional mean of the initial compressed coordinates given $U$. The matrix $J$ is
the Fisher information, at equal means, of a Gaussian ranking experiment: one observes only
the order of $n$ independent unit-variance Gaussian variables and estimates their means. A
translation common to all coordinates does not change the word and does not contribute to $C_U$.

For consecutive initial sites and a uniform initial word, take $U=W_0$. The constant is then
\begin{equation}\label{eq:Cn}
 C_n=\frac{(n-1)(2n+5)}{144}\sum_{k=1}^n(\E G_{(k)})^2,
\end{equation}
where $G_{(1)}<\cdots<G_{(n)}$ are the order statistics of $n$ independent standard normal
variables. In particular, $C_2=1/(8\pi)$ and
$C_3=11/(16\pi)$. In the same setting the information in the full configuration at time $t$
also decays as $t^{-1}$, and the word retains asymptotically
the fraction $\frac1n\sum_{k=1}^n(\E G_{(k)})^2$ of it.
This fraction equals $1/\pi$ for $n=2$, is nondecreasing in $n$, and tends to one as
$n\to\infty$.

The decay exponent depends on the exchange rule. In ordinary multispecies TASEP a
particle exchanges only with a weaker particle or a vacancy immediately to its right.
For two particles on fixed sites with both initial words of positive probability,
$I(W_0;W_t)$ then decays as $t^{-1/2}$ (Section~\ref{sec:two}). In both models the total variation
distance between the laws of $W_t$ under the two initial words decays as $t^{-1/2}$.
In this two-particle comparison the limiting law of $W_t$ under long-range swaps is uniform
and the information is quadratic in that distance. In ordinary TASEP the limiting law is
concentrated on one word and the information is linear in it.

\paragraph{Long times for three particles with a repeated species.}
The compositions $(2,1)$ and $(1,2)$ each admit three words and have
the same composition entropy
\begin{equation*}
H_{\rm comp}=-\sum_i(n_i/n)\log(n_i/n),
\end{equation*}
the entropy of the species label of a uniformly chosen particle.
Suppose that the particles initially occupy sites $0,1,2$ and that
$W_0$ is uniform on the three words of the chosen composition. Then
$H(W_0)=\log 3$ in both cases. For each composition, the two rules give the same limit of $tI(W_0;W_t)$
as $t\to\infty$. This limit is three times as large when the weaker
species occurs twice as when the stronger species occurs twice
(Theorem~\ref{thm:repeated}). The limiting word laws are nonuniform.
For either composition, the limiting probability of a word under the
drop--push-type rule equals the limiting probability of its reversal
under the TASEP-type rule.

\paragraph{Short times for arbitrary composition.}
Fix $n$ consecutive initial sites and choose $W_0$ uniformly from the
words with composition $\boldsymbol n=(n_1,\ldots,n_q)$. The entropy of
the initial word is $H(W_0)=S_{\rm ord}(\boldsymbol n)$, where
$S_{\rm ord}(\boldsymbol n):=\log(n!/\prod_{i=1}^q n_i!)$.
When all species follow the same rule, write
$I_{\boldsymbol n}^{\alpha}(t)=I(W_0;W_t)$, with $\alpha=0$ for the
TASEP-type rule and $\alpha=1$ for the drop--push-type rule. As
$t\downarrow0$,
\[
 I_{\boldsymbol n}^{\alpha}(t)
 =S_{\rm ord}(\boldsymbol n)
  -r_\alpha(\boldsymbol n)\,t\log(1/t)+O_{\boldsymbol n}(t).
\]
The coefficient $r_\alpha(\boldsymbol n)$ equals $n$ times the
probability that the first clock ring changes the species word
(Theorem~\ref{thm:composition}).

Splitting one species by replacing an entry $a+b$ of $\boldsymbol n$
with two adjacent positive entries $a,b$ strictly increases
$H_{\rm comp}$, $S_{\rm ord}$ and $r_\alpha$. Permuting the entries
of $\boldsymbol n$ preserves both entropies and the drop--push-type
coefficient $r_1$. The TASEP-type coefficient $r_0$ can depend on the
permutation and is largest when the multiplicities are nondecreasing
from weakest to strongest, that is, when $n_1\le\cdots\le n_q$
(Corollary~\ref{cor:diversity}).

\subsection{Relation to existing work}

The multispecies long-range swap model was introduced in
\cite{Lee2026}. For the drop--push-type rule, that paper establishes
Bethe ansatz integrability and gives a multiple contour integral for
the matrix of finite-time transition probabilities (Theorem~3.14).
A subsequent extension~\cite{LeeJSTAT2026} assigns each species its own parameter
for interpolating between the TASEP-type and drop--push-type rules
when two particles of that species meet. For a single species, a one-parameter family
interpolating between it and TASEP was solved in \cite{Ali}. For background on
exclusion processes, see Liggett~\cite[Chapter~VIII]{Liggett1985}. Tracy and
Widom~\cite{Tracy-Widom-2013} study nearest-neighbor multispecies
ASEP, and Lee~\cite{Lee-2021} treats multispecies TASEP with
species-dependent jump rates. Related models include the oriented swap process~\cite{AHR} and the
TASEP speed process~\cite{AAV}. For the oriented swap process,
Bufetov, Gorin and Romik~\cite[Theorem~1.2]{BGR} proved that the
centered and scaled absorption time converges to the GOE Tracy--Widom law.

Other multispecies models also admit reductions to fewer species. For the
multispecies $t$-PushTASEP on a finite ring, Ayyer, Martin
and Williams~\cite[Section~2.1]{AMW} show that merging species that
are adjacent in the priority order gives another $t$-PushTASEP on
the same ring. Our reduction is of a different kind. In the long-range swap
model on $\Z$, we subtract
from each particle's position the number of stronger particles to
its left. Conditional on the initial configuration, the resulting
coordinates evolve as independent single-species processes, one for
each species (Proposition~\ref{prop:blockreduction}).

Two works treat related information quantities in other settings. Wood, Blythe
and Evans~\cite{WBE} compute the second-order R\'enyi
entropy of the stationary configuration distribution of open-boundary
TASEP. Jeong, Dytso and Cardone~\cite{JDC} study the probability of
error in recovering the order of a data vector from observations
corrupted by additive Gaussian noise. Here we instead measure the mutual
information between an initial input and the species order at a later time.

Riffle shuffling provides an information-theoretic precedent. The works
\cite{TrefethenTrefethen,StarkGaneshOConnell} measure the randomization of a deck by entropy.
Starting from a fixed order of $n$ cards, \cite{StarkGaneshOConnell} studies the relative entropy
to the uniform law of the order after $k$ shuffles, which equals $\log(n!)-H(\Sigma_k)$ for the
permutation $\Sigma_k$ produced by the $k$ shuffles. If the initial order is instead uniform and
independent of $\Sigma_k$, the shuffled order is uniform for every $k$, and the same quantity is
the mutual information between the initial and the shuffled order. Those works follow the entropy as the number of shuffles and the deck size vary, while
our long-time limit is taken at a fixed particle number $n$.

The quadratic expansion of information at weak signal is classical \cite{PvdM,Sason}.
The constants $\E G_{(k)}$ in \eqref{eq:Cn} are the expected normal order statistics used
in normal-scores rank tests \cite{Hoeffding51,Terry,HajekSidak}. In
Theorem~\ref{thm:main} the model-specific steps
are the coordinate representation, the passage from the Poisson coordinates back to the
discrete word, and the evaluation of the coefficient. In Theorem~\ref{thm:repeated} the
classical two-particle transition laws of the single-species processes
\cite{Schutz-1997,Ali} give the marginal laws of the repeated pair. The new calculation
identifies which initial displacements survive the projection to the species word and
evaluates their information coefficients. At short times
(Theorem~\ref{thm:composition}) the new result is the
exact dependence of the loss coefficient on the composition and on the placement of
multiplicities.

\subsection{Organization of the paper}

Section~\ref{sec:model} defines the particle dynamics and the ordering
channel. For distinct species, Section~\ref{sec:coords} introduces
compressed coordinates with independent Poisson increments, and
Section~\ref{sec:info} uses these coordinates to derive the long-time
asymptotics of $I(U;W_t)$. Section~\ref{sec:two} compares the exact
two-particle ordering channels for long-range swaps and ordinary
multispecies TASEP and determines the information decay rate in each
model. Section~\ref{sec:types} studies short-time information loss for
arbitrary species compositions and long-time information decay for
the three-particle compositions $(2,1)$ and $(1,2)$.
Section~\ref{sec:discussion} concludes with a discussion and future problems.

All theorems are proved in the main text. Supplement~S1 provides
programs for reproducing the numerical results, including the data
plotted in Figures~\ref{fig:two} and~\ref{fig:late}. Generative AI tools were
used to verify proofs, to develop verification programs, and to produce figures and tables as described in
the Acknowledgments.

\section{The particle model and the ordering channel}\label{sec:model}

Consider $n\ge2$ particles on $\Z$, with at most one particle per site.
We label the species from $1$ (weakest) to $q$ (strongest). A
configuration is a map $\eta\colon\Z\to\{0,1,\ldots,q\}$ with exactly
$n$ nonzero entries. At an occupied site $x$, the label $\eta(x)$ gives
the species of the particle at $x$. A vacant site has label $0$.
Several particles may belong to the same species.

\begin{definition}[Long-range swap dynamics]\label{def:lr}
Each particle has an independent rate-one Poisson clock. When the clock
of a particle of species $i$ at site $x$ rings, its target is determined
from the configuration immediately before the ring.
\begin{itemize}
\item \emph{TASEP-type rule.} The target is the first site $y>x$ with
      $\eta(y)\le i$.
\item \emph{Drop--push-type rule.} The target is the first site $y>x$ with
      $\eta(y)<i$.
\end{itemize}
The labels at $x$ and $y$ are then exchanged. Thus the active particle
jumps to $y$ if $y$ is vacant and swaps positions with the particle at
$y$ if that particle is weaker. Under the TASEP-type rule, a target of
the same species leaves the configuration unchanged. In all cases,
the labels at sites other than $x$ and $y$ retain their values.
\end{definition}

Since only finitely many sites are occupied, there is always a vacancy
to the right of $x$, so the target exists under either rule. The total
clock rate is $n$, including rings that leave the configuration
unchanged. From here through Section~\ref{sec:two}, assume that each species
occurs exactly once, so $q=n$. The two rules of
Definition~\ref{def:lr} then give the same update at every clock ring.
Section~\ref{sec:types} returns to repeated species.

\begin{figure}[ht]
\centering
\begin{tikzpicture}[x=1.15cm,y=1.25cm,
  site/.style={draw=black!55,minimum size=7mm,inner sep=0pt,font=\small},
  s1/.style={site,fill=lrblue!12},
  s2/.style={site,fill=lrblue!30},
  s3/.style={site,fill=lrblue!55},
  lab/.style={font=\footnotesize}]
\foreach \x in {0,...,4}
  \node[lab,black!60] at (\x,0.62) {\x};
\node[lab,black!60,anchor=east] at (-0.55,0.62) {site};

% Initial configuration
\node[s1]   (a0) at (0,0) {1};
\node[s3]   (a1) at (1,0) {3};
\node[s2]   (a2) at (2,0) {2};
\node[site] (a3) at (3,0) {};
\node[site] (a4) at (4,0) {};
\node[lab,anchor=west] at (4.75,0) {$W=132,\quad T(r)=(0,1,1)$};
\draw[-{Stealth[length=2mm]},lrblue!70!black,thick]
  (a0.south) to[bend right=38]
  node[below,lab,pos=0.5] {particle $1$ jumps over $3$ and $2$}
  (a3.south);

% After the clock of species 1 rings
\node[site] (b0) at (0,-1.55) {};
\node[s3]   (b1) at (1,-1.55) {3};
\node[s2]   (b2) at (2,-1.55) {2};
\node[s1]   (b3) at (3,-1.55) {1};
\node[site] (b4) at (4,-1.55) {};
\node[lab,anchor=west] at (4.75,-1.55) {$W=321,\quad T(r)=(1,1,1)$};
\draw[{Stealth[length=2mm]}-{Stealth[length=2mm]},lrblue!70!black,thick]
  (b1.south) to[bend right=55]
  node[below,lab,pos=0.5] {particle $3$ initiates a swap with $2$}
  (b2.south);

% After the clock of species 3 rings
\node[site] (c0) at (0,-3.1) {};
\node[s2]   (c1) at (1,-3.1) {2};
\node[s3]   (c2) at (2,-3.1) {3};
\node[s1]   (c3) at (3,-3.1) {1};
\node[site] (c4) at (4,-3.1) {};
\node[lab,anchor=west] at (4.75,-3.1) {$W=231,\quad T(r)=(1,1,2)$};
\end{tikzpicture}
\caption{Two successive moves of the long-range swap dynamics for $n=3$.
The first move spans three sites, and the second exchanges two particles.
At each ring, the compressed coordinate
$T_i(r)=r_i-\#\{j>i:r_j<r_i\}$ of the active species $i$ increases by
one, and all other compressed coordinates stay fixed
(Theorem~\ref{thm:poisson}).}
\label{fig:moves}
\end{figure}
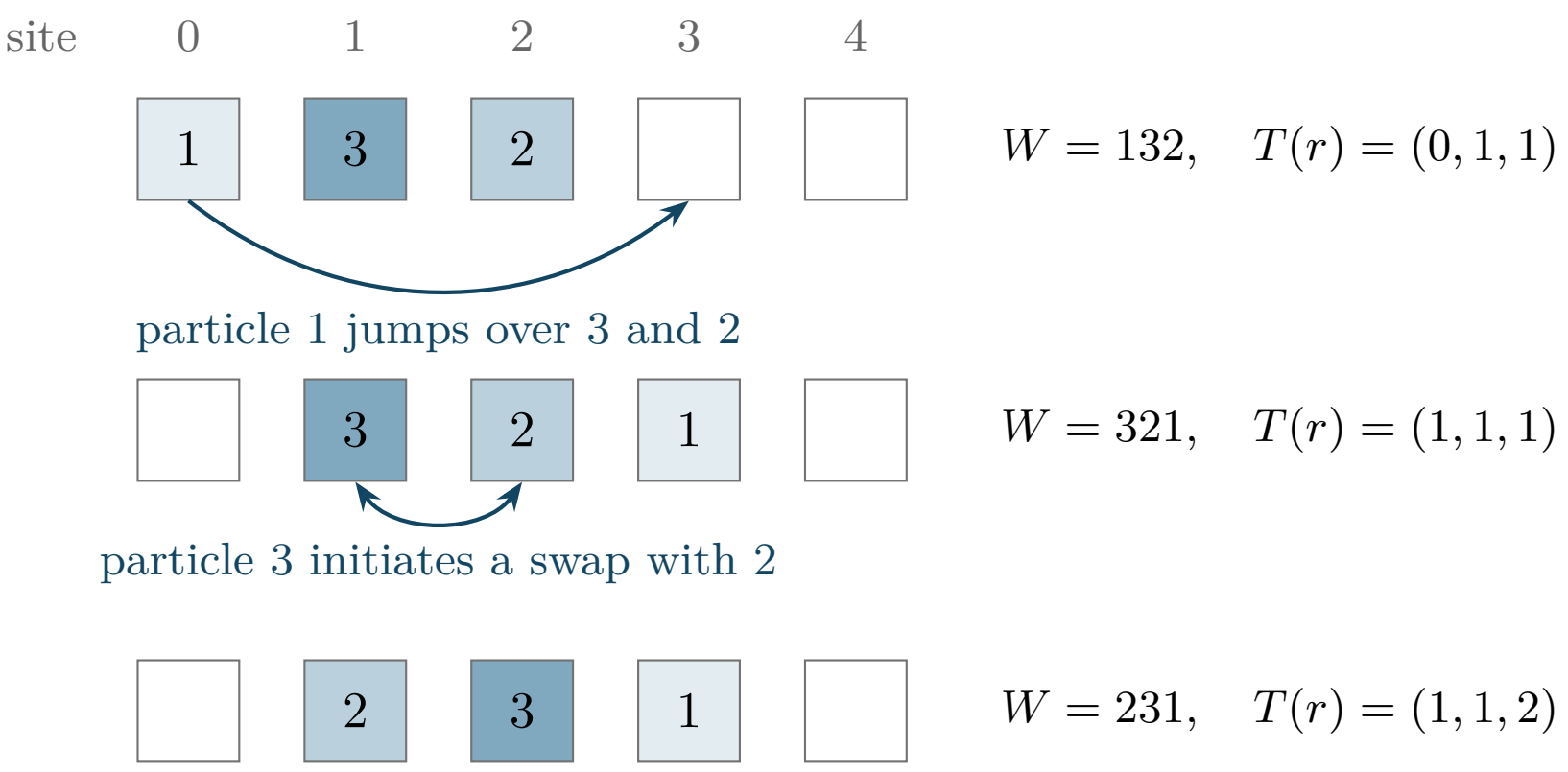

Write $r_i$ for the position of species $i$. A configuration is then
represented by
\[
 r=(r_1,\ldots,r_n)\in\X_n
 :=\{r\in\Z^n:r_i\ne r_j \text{ whenever } i\ne j\}.
\]
Let $S_n$ be the set of permutations of $\{1,\ldots,n\}$, written in
one-line notation. The species word $W(r)$ is the unique permutation
$\pi=(\pi_1,\ldots,\pi_n)$ satisfying
\[
 r_{\pi_1}<\cdots<r_{\pi_n}.
\]
Thus $W_t=W(r(t))$ lists the species from left to right at time $t$.
Figure~\ref{fig:moves} shows two successive moves.

\paragraph{Fixed initial sites.}
Fix sites $y_1<\cdots<y_n$ in $\Z$ and write $Y=(y_1,\ldots,y_n)$.
Draw the initial word $W_0$ from a prior
$\rho=(\rho_\nu)_{\nu\in S_n}$, independently of the clocks. For each
$\nu=(\nu_1,\ldots,\nu_n)\in S_n$, let $\PP_{Y,\nu}$ denote the law
of the process started with species $\nu_k$ at site $y_k$ for
$1\le k\le n$. The ordering channel $K_t$ and its output law $\mu_t$
are given by
\begin{equation}\label{eq:channel}
 K_t(\nu,\pi)=\PP_{Y,\nu}(W_t=\pi),\qquad
 \mu_t(\pi)=\PP(W_t=\pi)
           =\sum_{\nu\in S_n}\rho_\nu K_t(\nu,\pi).
\end{equation}
The starting configuration specifies the row $K_t(\nu,\cdot)$ for
every $\nu$, including those with $\rho_\nu=0$. When $\rho_\nu>0$,
this row is the conditional law of $W_t$ given $W_0=\nu$. The mutual
information defined in the introduction can therefore be written as
\[
 I(W_0;W_t)
 =H(W_t)-H(W_t\mid W_0)
 =H(\mu_t)-\sum_{\nu\in S_n}\rho_\nu H\bigl(K_t(\nu,\cdot)\bigr).
\]
We use the convention $0\log0=0$.

\paragraph{Initial distributions.}
More generally, let the initial configuration $r(0)$ have law $\lambda$
on $\X_n$, independently of the clocks, and take $U=f(r(0))$ as the
channel input. Write $\omega_u=\PP(U=u)$. For each $u$ with $\omega_u>0$,
define
\[
 K_t(u,\pi)=\PP(W_t=\pi\mid U=u),\qquad \pi\in S_n.
\]
This row averages the output laws from the configurations $r$ satisfying
$f(r)=u$, with weights $\PP(r(0)=r\mid U=u)$. The output law is
\[
 \mu_t(\pi)=\sum_{u:\,\omega_u>0}\omega_u K_t(u,\pi).
\]
For fixed initial sites $Y$ and $U=W_0$, the conditional rows agree
with those in \eqref{eq:channel} for every $\nu$ with $\rho_\nu>0$,
and the two formulas give the same output law. Taking $U=r(0)$ measures
the information that $W_t$ retains about the full initial configuration.

For probability laws $\mathsf P$ and $\mathsf Q$ on a common countable
alphabet $\mathcal A$, the relative entropy and the total variation
distance are
\[
 D(\mathsf P\|\mathsf Q)
 :=\sum_{\substack{x\in\mathcal A\\\mathsf P(x)>0}}
       \mathsf P(x)\log\frac{\mathsf P(x)}{\mathsf Q(x)},
 \qquad
 \|\mathsf P-\mathsf Q\|_{\TV}
 :=\frac12\sum_{x\in\mathcal A}|\mathsf P(x)-\mathsf Q(x)|.
\]
The relative entropy is $+\infty$ if $\mathsf P(x)>0$ and
$\mathsf Q(x)=0$ for some $x$. With this notation,
\[
\begin{aligned}
 I(U;W_t)
 &=\sum_{u:\,\omega_u>0}
   \omega_u D\bigl(K_t(u,\cdot)\,\|\,\mu_t\bigr)\\
 &=\sum_{u:\,\omega_u>0}\omega_u
   \sum_{\substack{\pi\in S_n\\K_t(u,\pi)>0}}
   K_t(u,\pi)\log\frac{K_t(u,\pi)}{\mu_t(\pi)}.
\end{aligned}
\]
Every term in the inner sum has
$\mu_t(\pi)\ge\omega_u K_t(u,\pi)>0$. Throughout, sums over conditional
input laws range over values $u$ with $\omega_u>0$. The long-time
results in Section~\ref{sec:info} assume that $\lambda$ has finite
support.

\begin{remark}[No autonomous Markov dynamics on words]\label{rem:semigroup}
Consider two particles in the order $12$. If they are adjacent, the
word changes to $21$ at instantaneous rate one. If there is a vacancy
between them, this rate is zero. Thus configurations with the same word
can have different rates of transition to another word. Even for fixed initial sites $Y$, the word channels $(K_t)_{t\ge0}$ in
\eqref{eq:channel} do not form a time-homogeneous Markov semigroup on
$S_n$. For any two distinct input words, the total variation distance
between their rows is of exact order $t^{-1/2}$ as $t\to\infty$
(Corollaries~\ref{cor:fixed} and~\ref{cor:tv}). In a finite-state
continuous-time Markov semigroup, any row distance that tends to zero
decays exponentially fast.

For $0\le s\le t$, the Markov property of the full configuration
process and the identity $W_t=W(r(t))$ give the Markov chain
\[
 W_0\longrightarrow r(s)\longrightarrow r(t)\longrightarrow W_t.
\]
The data-processing inequality therefore gives
\[
 I(W_0;W_t)\le I(W_0;r(t))\le I(W_0;r(s)).
\]
Comparing $I(W_0;W_t)$ with $I(W_0;W_s)$ requires a separate argument.
For two particles on fixed initial sites, Proposition~\ref{prop:tv2}
shows that $I(W_0;W_t)$ is nonincreasing in $t$ for every prior on
the initial word. For $n\ge3$ distinct species on fixed initial sites,
this monotonicity question remains open (Section~\ref{sec:discussion}).
\end{remark}

\section{Compression to Poisson coordinates}\label{sec:coords}

When a particle of species $i$ scans to the right, it passes every stronger particle and
stops at the first weaker particle or vacancy. Remove the sites occupied by species
$j>i$ and close the resulting gaps. Each deleted site to the left of $r_i$ reduces
its coordinate by one. For $r\in\X_n$, set

\begin{equation}\label{eq:T}
c_i(r)=\#\{j>i:r_j<r_i\},\qquad T_i(r)=r_i-c_i(r),\qquad T(r)=(T_1(r),\dots,T_n(r)).
\end{equation}
Each species uses its own reduced lattice, so distinct particles can have equal compressed
coordinates. In the top row of Figure~\ref{fig:moves}, species $1,3,2$ occupy physical
sites $0,1,2$, respectively, and site $3$ is vacant. For species $1$, deleting sites $1$
and $2$ makes site $3$ the next site after $0$. When species $1$ jumps from $0$ to $3$,
its compressed coordinate goes from $0$ to $3-2=1$.

\begin{theorem}[Poisson representation in compressed coordinates]\label{thm:poisson}
On the distinct-species sector the map $T:\X_n\to\Z^n$ is a bijection. If species $i$ activates, the compressed vector changes by $e_i$, the $i$-th standard basis vector. Consequently, with the same clocks as in Definition~\ref{def:lr},
\begin{equation}\label{eq:Z}
Z_i(t):=T_i(r(t))=T_i(r(0))+N_i(t),\qquad 1\le i\le n,
\end{equation}
where $N_1,\dots,N_n$ are independent rate-one Poisson processes. The inverse map $\Psi=T^{-1}$ satisfies
\begin{equation}\label{eq:Psibound}
0\le\Psi_i(z)-z_i\le n-i .
\end{equation}
\end{theorem}

\begin{proof}
For a finite set $S\subset\Z$ of occupied sites, define
\[
 F_S(x)=x-\#\{s\in S:s<x\},\qquad x\in\Z\setminus S.
\]
If $x<y$ are successive sites outside $S$, all $y-x-1$ sites between them lie in $S$.
Hence $F_S(y)-F_S(x)=1$. For nonempty $S$, we have $F_S(x)=x$ for $x<\min S$ and
$F_S(x)=x-|S|$ for $x>\max S$. Thus the successive values cover every integer exactly
once, and $F_S$ is a bijection from $\Z\setminus S$ onto $\Z$. For $S=\varnothing$, $F_S$ is the identity.

To prove that $T$ is onto, fix $z\in\Z^n$. Place species $n,n-1,\ldots,1$ in that
order. When the stronger species occupy $S$, place species $i$ at $F_S^{-1}(z_i)$.
Inserting weaker particles later does not change $T_i$, so the resulting configuration
$r$ satisfies $T(r)=z$. To prove injectivity, suppose $T(r)=T(r')=z$. The strongest
species is at $z_n$ in both configurations. If all stronger species have the same sites
in $r$ and $r'$, the same set $S$ determines $r_i=r'_i=F_S^{-1}(z_i)$. Descending
induction gives $r=r'$. Finally, $\Psi_i(z)-z_i=c_i(\Psi(z))$ counts at most $n-i$
stronger particles, which proves \eqref{eq:Psibound}.

Now consider an activation of species $i$ at $x$, with target $y=x+\ell$. The $\ell-1$
intervening particles are all stronger than $i$. If $y$ is vacant, $r_i$ increases by
$\ell$, while $c_i$ increases by $\ell-1$; hence $T_i$ increases by one. An intervening
particle is stronger than $i$, so $i$ does not enter its count. A particle outside
$[x,y]$ stays on the same side of $i$. Hence all other compressed coordinates are unchanged.

If the target contains species $j<i$, the same calculation gives an increment of one for $T_i$. Species $j$ moves to the left by $\ell$. Its count $c_j$ decreases by $\ell$, since it crosses species $i$ and the $\ell-1$ intervening stronger particles. Thus $T_j$ is unchanged. For a third particle strictly between $x$ and $y$,
its species $k$ satisfies $k>i>j$. Neither moving particle enters its count
$c_k$, and its site is fixed. For a particle outside $[x,y]$, both exchanged
particles remain on the same side of it, so its count and site are unchanged.

With distinct species every ring produces a jump or a swap: a target $y>x$ exists
because there are finitely many particles, and its label differs from $i$.
Consequently the compressed coordinate of species $i$ increases by one at
each ring of its own clock and at no other time. Thus \eqref{eq:Z} holds with
$N_i(t)$ the number of rings of the clock of species $i$ up to time $t$. The
clocks are independent and have rate one, so $N_1,\dots,N_n$ are independent
rate-one Poisson processes.
\end{proof}

Conditional on the initial configuration, the compressed coordinate processes are
independent. For a random initial configuration, their starting values can be dependent,
and this dependence can persist in their unconditional joint law.

\begin{lemma}[Translation covariance]\label{lem:translation}
Let $\one=(1,\ldots,1)\in\Z^n$. For $c\in\Z$ and $r\in\X_n$, $T(r+c\one)=T(r)+c\one$ and $W(r+c\one)=W(r)$. Hence $\Psi(z+c\one)=\Psi(z)+c\one$ for $z\in\Z^n$.
\end{lemma}

\begin{proof}
The counts $c_i(r)$ and the word $W(r)$ depend only on the relative order of the positions.
\end{proof}

\begin{corollary}[Full transition kernel]\label{cor:kernel}
For $r,r'\in\X_n$, write $\Delta_i=T_i(r')-T_i(r)$. Then
\begin{equation}\label{eq:kernel}
\PP(r(t)=r'\mid r(0)=r)=
\begin{cases}
\displaystyle e^{-nt}\prod_{i=1}^n\frac{t^{\Delta_i}}{\Delta_i!}, & \Delta_i\ge0\text{ for every }i,\\[2ex]
0, & \text{otherwise.}
\end{cases}
\end{equation}
\end{corollary}

\begin{proof}
Apply the bijection $T$ and the independent Poisson increment law. The bijection also gives normalization over all final configurations.
\end{proof}

\begin{remark}[Relation to the Bethe-ansatz kernel]\label{rem:bethe}
Corollary~\ref{cor:kernel} also follows from the matrix-valued contour integral
of \cite[Theorem~3.14, Eq.~(65)]{Lee2026}. For distinct species, the scattering matrix
of \cite[Eq.~(57)]{Lee2026} exchanges each pair of labels $a<b$, with weight $\xi_a^{-1}$
when $a$ moves to the right of $b$ and $\xi_a$ when it moves to the left, where $\xi_i$ is
the spectral variable carried by species $i$. Hence a single permutation contributes to
the entry for the words of $r$ and $r'$, with amplitude $\prod_i\xi_i^{c_i(r)-c_i(r')}$.
Together with the position factor $\prod_i\xi_i^{r'_i-r_i-1}$, the integral becomes the product of
\[
 \frac1{2\pi\mathrm i}\oint_{|\xi|=R}\xi^{\Delta_i-1}
             e^{t(\xi^{-1}-1)}\,d\xi,\qquad R>1 ,
\]
and each factor equals $e^{-t}t^{\Delta_i}/\Delta_i!$ if $\Delta_i\ge0$ and zero otherwise.
\end{remark}

Summing \eqref{eq:kernel} over final configurations with a given word
gives the ordering channel. For fixed initial sites $Y$ and word
$\nu$, let $r^{Y,\nu}$ be the configuration with species $\nu_k$
at $y_k$. Set
\begin{equation}\label{eq:mE}
 m^\nu=T(r^{Y,\nu}),\qquad
 \mathcal C_\pi=\{z\in\Z^n:W(\Psi(z))=\pi\},\qquad \pi\in S_n.
\end{equation}
The sets $\mathcal C_\pi$ partition $\Z^n$, and each is invariant under
$z\mapsto z+c\one$ for $c\in\Z$ by Lemma~\ref{lem:translation}.
Writing $N(t)=(N_1(t),\ldots,N_n(t))$, we obtain from \eqref{eq:Z}
\begin{equation}\label{eq:Kpoisson}
 K_t(\nu,\pi)
 =\PP\bigl(m^\nu+N(t)\in\mathcal C_\pi\bigr)
 =e^{-nt}\sum_{\substack{k\in\Z_{\ge0}^n\\m^\nu+k\in\mathcal C_\pi}}
   \prod_{i=1}^n\frac{t^{k_i}}{k_i!}.
\end{equation}

For finitely supported $\lambda$, \eqref{eq:Kexp} and \eqref{eq:rankrate}
in Section~\ref{sec:info} give $K_t(u,\pi)=1/n!+O_{n,\lambda}(t^{-1/2})$.
For arbitrary $\lambda$ and any $u$ with $\PP(U=u)>0$, averaging the
fixed-configuration limits over $r(0)$ given $U=u$ yields
$K_t(u,\pi)\to 1/n!$ by bounded convergence.

\begin{remark}[Species-dependent clock rates]\label{rem:variations}
Species-dependent jump rates for nearest-neighbor multispecies TASEP
were studied in \cite{Lee-2021}. If the clock of species $i$ has rate $b_i>0$,
the identity \eqref{eq:Z} still holds, with $N_i$ a Poisson process of rate $b_i$.
For pairwise distinct rates, $r_i(t)/t\to b_i$ almost surely, so $W_t$ is eventually
the word $w_b$ that lists the species in increasing order of their rates. For a
finitely supported initial law, a Chernoff bound gives $\PP(W_t\ne w_b)=O(e^{-ct})$
for some $c>0$, and $I(W_0;W_t)\le H(W_t)$ is then exponentially small in $t$.
The following sections use rate-one clocks.
\end{remark}

\section{The asymptotic information coefficient}\label{sec:info}

Let $G_1,\dots,G_n$ be independent standard normals and let $G_{(1)}<\dots<G_{(n)}$ denote their order statistics. Set
\begin{equation}\label{eq:aAP}
a_k=\E G_{(k)},\qquad A_n=\sum_{k=1}^na_k^2,\qquad P=\Id-\frac1n\one\one^{\T}.
\end{equation}
In particular, $\sum_ka_k=0$. The matrix $P$ is the orthogonal projection onto the subspace of vectors whose coordinates sum to zero. Throughout, $\phi$ and $\Phi$ denote the standard normal density and distribution function.

\begin{theorem}[Information decay for fixed $n$]\label{thm:main}
Fix $n\ge2$ distinct species. Let $r(0)$ have a law $\lambda$ with finite support in $\X_n$, independent of the clocks, and let $U=f(r(0))$ be any function of the initial configuration. Put $M=T(r(0))$, $m_u=\E[M\mid U=u]$ and $\bar m=\E M$. As $t\to\infty$,
\begin{equation}\label{eq:main}
I(U;W_t)=\frac{C_U}{t}+O_{n,\lambda}(t^{-3/2}),\qquad C_U=\frac{A_n}{2(n-1)}\,\E\bigl\|P(m_U-\bar m)\bigr\|^2 .
\end{equation}
We have $C_U=0$ exactly when $Pm_U$ is almost surely constant.
For such inputs $I(U;W_t)=O_{n,\lambda}(t^{-2})$. If $C_U>0$,
$I(U;W_t)\sim C_U/t$.
\end{theorem}

If $U'$ is a function of $U$, conditional Jensen's inequality gives
$C_{U'}\le C_U$, in agreement with the data-processing inequality.
We now specialize to fixed initial sites and $U=W_0$.

\begin{corollary}[Fixed sites]\label{cor:fixed}
Fix the initial occupied sites $Y=(y_1,\ldots,y_n)$ with
$y_1<\cdots<y_n$, and let the initial word $W_0$ have law
$\rho=(\rho_\nu)_{\nu\in S_n}$, independently of the clocks.
With $m^\nu$ as in \eqref{eq:mE}, put
$\bar m=\sum_{\nu\in S_n}\rho_\nu m^\nu$. As $t\to\infty$,
\begin{equation}\label{eq:fixed}
 I(W_0;W_t)=\frac{C(Y,\rho)}{t}+O_{n,Y,\rho}(t^{-3/2}),
 \qquad
 C(Y,\rho)=\frac{A_n}{2(n-1)}
 \sum_{\nu\in S_n}\rho_\nu\bigl\|P(m^\nu-\bar m)\bigr\|^2.
\end{equation}
The coefficient $C(Y,\rho)$ is positive if and only if $\rho$ assigns
positive probability to at least two words. If $\rho$ is a point mass,
then $I(W_0;W_t)=0$ for every $t\ge0$. More generally, if $U=r(0)$ in Theorem~\ref{thm:main}, then $C_U=0$
if and only if the support of $\lambda$ is contained in
$\{r_\star+c\one:c\in\Z\}$ for some $r_\star\in\X_n$.
\end{corollary}

\subsection{A likelihood expansion}

\begin{lemma}\label{lem:likelihood}
Let $t>0$, and let $N=(N_1,\ldots,N_n)$ have independent $\Pois(t)$
coordinates. Denote its law by $Q_t$, and set $V_i=(N_i-t)/\sqrt t$
and $V=(V_1,\ldots,V_n)$. For fixed $m\in\Z_{\ge0}^n$, the law of
$m+N$ has density $L_m$ with respect to $Q_t$, given by
\begin{equation}\label{eq:L}
 L_m(z)=\prod_{i=1}^n\frac{(z_i)_{m_i}}{t^{m_i}},
 \qquad z\in\Z_{\ge0}^n,
\end{equation}
where $(z)_a=z(z-1)\cdots(z-a+1)$ for integers $a\ge1$ and $(z)_0=1$.
The remainder in the expansion
\begin{equation}\label{eq:Lexp}
 L_m(N)=1+t^{-1/2}\,m\cdot V+\operatorname{rem}_{m,t}
\end{equation}
satisfies, for every $1\le p<\infty$,
\[
 \|\operatorname{rem}_{m,t}\|_{L^p(Q_t)}
 =O_{m,p}(t^{-1})\qquad\text{as }t\to\infty.
\]
\end{lemma}

\begin{proof}
For $z_i\ge m_i$, the Poisson mass function gives
\[
 \frac{\PP(m_i+N_i=z_i)}{\PP(N_i=z_i)}
 =\frac{z_i!}{t^{m_i}(z_i-m_i)!}
 =\frac{(z_i)_{m_i}}{t^{m_i}}.
\]
For $0\le z_i<m_i$, both the shifted mass and the falling factorial
vanish. Independence now gives \eqref{eq:L}. Evaluating at $N$, we obtain
\[
 L_m(N)=\prod_{i=1}^n\prod_{j=0}^{m_i-1}
 \left(1+\frac{V_i}{\sqrt t}-\frac jt\right).
\]
The terms obtained by selecting $V_i/\sqrt t$ from one factor and $1$
from all the others sum to $t^{-1/2}m\cdot V$. After subtracting
$1+t^{-1/2}m\cdot V$, the remaining terms are monomials in the
coordinates of $V$ with coefficients $O_m(t^{-1})$.

For each integer $k\ge1$, the centered moment $\E[(N_i-t)^{2k}]$ is
a polynomial in $t$ of degree at most $k$, as follows by expanding
$\E[e^{s(N_i-t)}]=\exp\{t(e^s-1-s)\}$ at $s=0$. Thus every fixed
absolute moment of $V_i$ is uniformly bounded for $t\ge1$.
H\"older's inequality bounds the $L^p(Q_t)$ norm of each monomial
uniformly in $t\ge1$. Summing the finitely many terms gives the
asserted remainder bound.
\end{proof}

\begin{lemma}[Permutation second moment]\label{lem:perm}
Let $n\ge2$. For $\pi\in S_n$, let $\rank_\pi(i)$ be the position of
species $i$ in $\pi$, and define $a^\pi\in\R^n$ by
$a_i^\pi=a_{\rank_\pi(i)}$. Then
\begin{equation}\label{eq:perm}
 \frac1{n!}\sum_{\pi\in S_n}(v\cdot a^\pi)^2
 =\frac{A_n}{n-1}\|Pv\|^2,
 \qquad v\in\R^n.
\end{equation}
\end{lemma}

\begin{proof}
Let $\pi$ be uniform on $S_n$. Each rank $\rank_\pi(i)$ is uniform on
$\{1,\ldots,n\}$, so $\E[a_i^\pi]=0$ and $\E[(a_i^\pi)^2]=A_n/n$.
For $i\ne j$, the two ranks are uniform over the $n(n-1)$ ordered
pairs of distinct ranks. Hence
\[
 \E[a_i^\pi a_j^\pi]
 =\frac{\sum_{k\ne\ell}a_ka_\ell}{n(n-1)}
 =-\frac{A_n}{n(n-1)},
\]
where the last equality uses $\sum_{k=1}^n a_k=0$.
It follows that $\E[a^\pi(a^\pi)^{\T}]=A_nP/(n-1)$.
Taking the quadratic form in $v$ and using
$v^{\T}Pv=\|Pv\|^2$ gives \eqref{eq:perm}.
\end{proof}

Since $\E[v\cdot a^\pi]=0$ for uniform $\pi$, \eqref{eq:perm} is the
classical variance formula for a linear rank statistic with scores
$a_k$ and regression constants $v_i$~\cite{HajekSidak}.

\subsection{Proof of Theorem~\ref{thm:main} and its corollaries}

\begin{proof}[Proof of Theorem~\ref{thm:main}]
Under a common change of origin $r'=r+c\one$, retain the same input
by setting $f_c(r')=f(r'-c\one)$. Lemma~\ref{lem:translation} preserves the
word, and both conditional and unconditional mean coordinates shift by
$c\one$. Thus the channel and $m_u-\bar m$ are unchanged. Since the support is finite, we may therefore assume that $T(r_\star)$ has nonnegative coordinates for every $r_\star$ in the support. (For fixed sites it suffices to translate so that $y_1=0$: if species $i$ initially occupies $y_k$, then $y_k\ge k-1$ and $c_i\le k-1$.)

Let $Q_t,N,V$ be as in Lemma~\ref{lem:likelihood}. Define
\[
b_\pi(t)=Q_t(\mathcal C_\pi),\qquad \beta_\pi(t)=\E_{Q_t}\bigl[V\one\{N\in \mathcal C_\pi\}\bigr].
\]
For a fixed configuration $r_\star$ with $m=T(r_\star)$, Theorem~\ref{thm:poisson} and the likelihood expansion, followed by Cauchy--Schwarz, give
\[
\PP(W_t=\pi\mid r(0)=r_\star)=\E_{Q_t}\bigl[L_m(N)\one\{N\in \mathcal C_\pi\}\bigr]=b_\pi(t)+t^{-1/2}\,m\cdot \beta_\pi(t)+O(t^{-1}).
\]
Averaging over the conditional law of $r(0)$ given $U=u$, and then over $U$,
\begin{equation}\label{eq:Kexp}
\begin{aligned}
K_t(u,\pi)&=b_\pi(t)+t^{-1/2}\,m_u\cdot \beta_\pi(t)+O(t^{-1}),\\
\mu_t(\pi)&=b_\pi(t)+t^{-1/2}\,\bar m\cdot \beta_\pi(t)+O(t^{-1}).
\end{aligned}
\end{equation}
All constants can be chosen uniformly over the finitely many configurations.

Next we evaluate $b_\pi$ and $\beta_\pi$. Break ties among $N_1,\ldots,N_n$ with independent continuous uniform
variables, and denote the resulting permutation by $\Pi$. The pairs
consisting of a count and its tie breaker are exchangeable. Each of
the $n!$ rankings therefore has probability $1/n!$, also conditionally
on the sorted counts. If $i$ has rank $k$ in $\pi$, it follows that
\[
 \E[V_i\one\{\Pi=\pi\}]=\E V_{(k)}/n!.
\]
The word $W(\Psi(N))$ can differ from $\Pi$ only if
$|N_j-N_l|\le n$ for a pair $j\ne l$, since
$0\le\Psi_i(N)-N_i\le n-i$. For any such pair and a fixed $i$ choose
one of $j,l$ different from $i$ and condition on all counts except the
chosen one. At most $2n+1$ integers qualify, and the largest Poisson mass
is $O(t^{-1/2})$ by Stirling's formula at its mode. The variable $|V_i|$
is measurable under this conditioning and $\E|V_i|\le1$. A union bound
therefore bounds both the mismatch probability and its expectation
weighted by $|V_i|$ by $O_n(t^{-1/2})$.

For $t\ge2$ write a $\Pois(t)$ variable as the sum of $k_t=\lfloor t\rfloor$
independent $\Pois(t/k_t)$ variables. For each centered summand divided by
$\sqrt t$ the third absolute moment is $O(t^{-3/2})$, uniformly since
$1\le t/k_t\le2$. Their variances sum to one. The Wasserstein normal
approximation for independent summands \cite[Eq.~(10)]{ChenStein2021}
gives a one-dimensional $1$-Wasserstein distance $O(t^{-1/2})$.
Couple the $n$ coordinates independently with standard normals using
one-dimensional quantile coupling. Sorting satisfies
$|v_{(k)}-g_{(k)}|\le\max_i|v_i-g_i|\le\sum_i|v_i-g_i|$, so
$|\E V_{(k)}-a_k|=O_n(t^{-1/2})$. Combining the two comparisons gives
\begin{equation}\label{eq:rankrate}
 b_\pi(t)=1/n!+O_n(t^{-1/2}),\qquad
 \beta_{\pi,i}(t)=a_i^\pi/n!+O_n(t^{-1/2}).
\end{equation}

Subtract the two expansions in \eqref{eq:Kexp}. This cancels the entire common term $b_\pi(t)$, including any lattice correction it contains. Using \eqref{eq:rankrate}, we obtain
\begin{equation}\label{eq:delta}
\delta_{u,\pi}(t):=K_t(u,\pi)-\mu_t(\pi)=\frac{(m_u-\bar m)\cdot a^\pi}{n!\sqrt t}+O_{n,\lambda}(t^{-1}).
\end{equation}
By \eqref{eq:Kexp} and \eqref{eq:rankrate}, $\mu_t(\pi)=1/n!+O(t^{-1/2})$,
so all denominators stay bounded away from zero. Write $x=\delta_{u,\pi}/\mu_t(\pi)$
and expand $(1+x)\log(1+x)=x+x^2/2+O(|x|^3)$.
Summing cancels the linear term because $\sum_\pi\delta_{u,\pi}=0$.
There are finitely many outputs and each $\delta_{u,\pi}=O(t^{-1/2})$. Thus
\[
 D(K_t(u,\cdot)\|\mu_t)
 =\tfrac12\sum_\pi\delta_{u,\pi}^2/\mu_t(\pi)+O(t^{-3/2}).
\]
Squaring \eqref{eq:delta} introduces an $O(t^{-3/2})$ error, and
$1/\mu_t(\pi)=n!+O(t^{-1/2})$. Averaging with the input probabilities
$\omega_u$ gives
\[
 I(U;W_t)=\frac1{2t}\sum_u\omega_u\,\frac1{n!}
 \sum_\pi[(m_u-\bar m)\cdot a^\pi]^2+O_{n,\lambda}(t^{-3/2}).
\]
Lemma~\ref{lem:perm} proves \eqref{eq:main}. Its coefficient is zero exactly
when $P(m_u-\bar m)=0$ for every $u$ of positive probability. Then
\eqref{eq:delta} gives $\delta_{u,\pi}=O(t^{-1})$ for every input and
output. The same Taylor expansion bounds the information by $O(t^{-2})$.
\end{proof}

\begin{proof}[Proof of Corollary~\ref{cor:fixed}]
Formula \eqref{eq:fixed} is \eqref{eq:main} with $U=W_0$ and $m_u=m^\nu$. Suppose that $P(m^\nu-m^{\nu'})=0$ for two words $\nu,\nu'$. Then $m^{\nu'}=m^\nu+c\one$ for an integer $c$, and by Lemma~\ref{lem:translation} and the injectivity of $T$ the initial configuration of $\nu'$ is the translate by $c$ of the initial configuration of $\nu$. Both have the finite occupied set $Y$, so $c=0$ and $\nu=\nu'$. Hence $C(Y,\rho)=0$ forces $\rho$ to be a point mass. The same argument proves the statement for $U=r(0)$; conversely, if $\lambda$ is carried by translates of one configuration, then $W_t$ is independent of $r(0)$ by Lemma~\ref{lem:translation}.
\end{proof}

The same expansion controls the distinguishability of two inputs directly.

\begin{corollary}[Total variation between rows]\label{cor:tv}
In the setting of Theorem~\ref{thm:main}, let $u,u'$ be labels of positive probability. Then
\begin{equation}\label{eq:tv}
\bigl\|K_t(u,\cdot)-K_t(u',\cdot)\bigr\|_{\TV}=\frac{1}{2\,n!\,\sqrt t}\sum_{\pi\in S_n}\bigl|(m_u-m_{u'})\cdot a^\pi\bigr|+O_{n,\lambda}(t^{-1}),
\end{equation}
and the leading coefficient is positive if and only if $P(m_u-m_{u'})\ne0$. Equivalently, an observer who knows that $U\in\{u,u'\}$ with equal prior probabilities identifies $U$ from $W_t$ with optimal success probability $\frac12+\frac12\|K_t(u,\cdot)-K_t(u',\cdot)\|_{\TV}$.
\end{corollary}

\begin{proof}
Subtract \eqref{eq:Kexp} for $u$ and $u'$ and use \eqref{eq:rankrate} and $\bigl||x|-|y|\bigr|\le|x-y|$. Positivity follows from Lemma~\ref{lem:perm}. The last statement is the standard expression of the Bayes risk for two hypotheses.
\end{proof}

\begin{remark}[Rate of convergence]\label{rem:rate}
The remainder in \eqref{eq:main} bounds the absolute error, and its sign can depend on
time. For $Y=(0,3,9)$ and the prior with $\rho_{123}=0.2$ and $\rho_{132}=0.8$, for example,
$tI(W_0;W_t)$ lies below $C(Y,\rho)\approx3.4759$ at $t=10^3$ and above it at
$t=1.6\cdot10^4$. The coefficient of a possible $t^{-3/2}$ correction remains undetermined.
\end{remark}

\subsection{Consecutive initial sites and a uniform prior}

\begin{corollary}\label{cor:Cn}
Suppose $Y=(0,1,\dots,n-1)$ and $\rho$ is uniform on $S_n$. Then $C(Y,\rho)=C_n$, with $C_n$ given by \eqref{eq:Cn}. In particular,
\begin{equation}\label{eq:C23}
C_2=\frac1{8\pi},\qquad C_3=\frac{11}{16\pi}.
\end{equation}
\end{corollary}

\begin{proof}
If species $i$ is in position $k$ of the initial word, then
\[
m_i=(k-1)-\#\{\text{larger labels before }i\}=\#\{\text{smaller labels before }i\}.
\]
A uniform permutation is constructed by inserting label $i$ uniformly into one of the $i$ positions in the word on labels $1,\dots,i-1$. These insertion choices are independent, and later insertions do not change the number of smaller labels preceding $i$. Thus $m_1,\dots,m_n$ are independent and
\[
m_i\sim\mathrm{Unif}\{0,\dots,i-1\},\qquad\Var(m_i)=\frac{i^2-1}{12}.
\]
It follows that
\[
\E\|P(m-\E m)\|^2=\tr(P\Cov(m))=\frac{n-1}{n}\sum_{i=1}^n\frac{i^2-1}{12}=\frac{(n-1)^2(2n+5)}{72}.
\]
Substituting this into \eqref{eq:fixed} proves \eqref{eq:Cn}. For $n=2$, the mean Gaussian order statistics are $(-1/\sqrt\pi,1/\sqrt\pi)$. For $n=3$, they are $(-3/(2\sqrt\pi),0,3/(2\sqrt\pi))$. These values yield \eqref{eq:C23}.
\end{proof}

\begin{lemma}[Large $n$]\label{lem:An}
For $n\ge2$, $A_n<n-1$, and $A_n/n$ is nondecreasing in $n$ with limit one. Consequently $C_n\sim n^3/72$.
\end{lemma}

The limit $A_n/n\to1$ is the Gaussian second-moment case of a theorem of Hoeffding \cite{Hoeffding53}. The martingale argument below also gives the monotonicity in $n$; the finite-$n$ bound follows from a separate variance identity.

\begin{proof}
Take an infinite sequence of independent standard normals $G_1,G_2,\dots$. For $j\ge2$ let $B_j=\one\{G_j<G_1\}$, set $\mathcal F_n=\sigma(B_2,\dots,B_n)$, and write $J_n=1+\sum_{j=2}^n B_j$. Conditional on a specified pattern of the $B_j$ with $J_n=k$, the density of $G_1$ is proportional to
\[
\phi(x)\Phi(x)^{k-1}(1-\Phi(x))^{n-k}.
\]
This is the density of the $k$th order statistic of $n$ standard normals. Thus, writing $a_k^{(n)}$ to display the sample size,
\[
\mathcal M_n:=\E[G_1\mid\mathcal F_n]=a_{J_n}^{(n)},\qquad
\E \mathcal M_n^2=\frac1n\sum_{k=1}^n(a_k^{(n)})^2=\frac{A_n}{n},
\]
because $J_n$ is uniform on $\{1,\dots,n\}$. The conditional expectations $\mathcal M_n$ form an $L^2$-bounded martingale, so their second moments are nondecreasing. Given $G_1$, the variables $B_j$ are independent Bernoulli variables with mean $\Phi(G_1)$. The conditional strong law recovers $\Phi(G_1)$, and hence $G_1$, from the infinite sequence $(B_j)_{j\ge2}$. Therefore $\mathcal M_n\to G_1$ in $L^2$ by martingale convergence, proving $A_n/n\to1$. For the finite-$n$ bound, let $\bar G=n^{-1}\sum_iG_i$.
Sorting preserves the sum of squared deviations from $\bar G$, so
\[
 n-1=\E\sum_{i=1}^n(G_i-\bar G)^2
 =A_n+\sum_{k=1}^n\Var(G_{(k)}-\bar G).
\]
For $n\ge2$ the maximum minus the sample mean is nonconstant, since the joint Gaussian
density is positive on every open ball. Thus $A_n<n-1$, and the asymptotic for $C_n$ follows from \eqref{eq:Cn}.
\end{proof}

Table~\ref{tab:Cn} lists the first values. Figure~\ref{fig:late}(a) compares $tI(W_0;W_t)$
with the predicted coefficients for $n=3,4$.

\begin{table}[ht]
\centering\small
\caption{The coefficient $C_n$ of \eqref{eq:Cn} and the ordering efficiency $A_n/n$ of Corollary~\ref{cor:efficiency}, from numerical integration of $\E G_{(k)}$.}\label{tab:Cn}
\begin{tabular}{@{}rrrrr@{}}
\toprule
$n$ & $A_n$ & $C_n$ & $C_n/(n^3/72)$ & $A_n/n$\\
\midrule
2 & 0.636620 & 0.039789 & 0.358 & 0.3183\\
3 & 1.432394 & 0.218838 & 0.584 & 0.4775\\
4 & 2.295659 & 0.621741 & 0.699 & 0.5739\\
5 & 3.195060 & 1.331275 & 0.767 & 0.6390\\
6 & 4.116565 & 2.429917 & 0.810 & 0.6861\\
8 & 5.999494 & 6.124483 & 0.861 & 0.7499\\
10 & 7.914272 & 12.366050 & 0.890 & 0.7914\\
12 & 9.848116 & 21.816312 & 0.909 & 0.8207\\
\bottomrule
\end{tabular}
\end{table}

\begin{remark}[Order of limits]\label{rem:largen}
Each $C_n$ is defined by first taking $t\to\infty$ at fixed $n$. The remainder in
Theorem~\ref{thm:main} has no established bound uniform in $n$.
\end{remark}

\subsection{Gaussian ranking and Fisher information}\label{sec:fisher}

The coefficient in Theorem~\ref{thm:main} can be identified from a Gaussian ranking experiment. For $g\in\R^n$ with distinct coordinates, let $W(g)$ be the word listing the indices of the coordinates of $g$ in increasing order.

\begin{proposition}[Gaussian ranking channel]\label{prop:ranking}
Let $M$ be a random vector taking finitely many values in $\R^n$, independent of $G=(G_1,\dots,G_n)$, and observe only the coordinate ordering $\Pi_\varepsilon=W(G+\varepsilon M)$. The Fisher information matrix for the ordering of $G+\vartheta$, evaluated at $\vartheta=0$, is
\begin{equation}\label{eq:Jrank}
J_{\rank}=\frac{A_n}{n-1}P .
\end{equation}
Moreover,
\begin{equation}\label{eq:weak}
I(M;\Pi_\varepsilon)=\frac{\varepsilon^2}2\tr\bigl(J_{\rank}\Cov(M)\bigr)+o(\varepsilon^2).
\end{equation}
\end{proposition}

\begin{proof}
Let $p_\pi(\vartheta)=\PP(W(G+\vartheta)=\pi)$. Differentiating the Gaussian density under its integral over the ordering cone gives
\[
p_\pi(0)=\frac1{n!},\qquad\nabla p_\pi(0)=\E\bigl[G\one\{W(G)=\pi\}\bigr]=\frac{a^\pi}{n!}.
\]
The last equality holds because the ranking of $G$ is independent of its order statistics, and on $\{W(G)=\pi\}$ the coordinate $G_i$ is the order statistic of rank $\rank_\pi(i)$. The score vector is therefore $a^\pi$, and averaging its outer product gives \eqref{eq:Jrank} by Lemma~\ref{lem:perm}. For \eqref{eq:weak}, expand $p_\pi(\varepsilon m)$ about zero and subtract its average over $M$. The difference is $\varepsilon(m-\E M)\cdot a^\pi/n!+O(\varepsilon^2)$. The finite-alphabet relative-entropy expansion used above proves \eqref{eq:weak}.
\end{proof}

The scores $a^\pi$ are the normal scores of rank-test theory: by Hoeffding's formula for rank probabilities under location alternatives \cite{Hoeffding51}, for a fixed direction $m$ with $Pm\ne0$ the locally most powerful rank test against the one-sided Gaussian shift alternatives $\varepsilon m$, $\varepsilon\downarrow0$, rejects for large values of the linear statistic $m\cdot a^\pi=\sum_im_i\,\E G_{(\rank_\pi(i))}$ \cite{HajekSidak,Terry}. By \eqref{eq:delta}, the coefficient of $t^{-1/2}$ in $K_t(u,\pi)-\mu_t(\pi)$ depends on $u$ only through $(m_u-\bar m)\cdot a^\pi$, and $C_U$ is half the variance of this statistic over a uniform permutation $\pi$, averaged over $u$. Theorem~\ref{thm:main} thus shows that the coefficient of the Gaussian experiment with input $m_U$ and $\varepsilon=t^{-1/2}$ survives the discrete Poisson dynamics and the physical ordering map. The effective input is the conditional mean $m_U$, not the compressed vector $M$ itself. By \eqref{eq:Jrank} and Lemma~\ref{lem:An}, $J_{\rank}\le P$ in the Loewner order, with strict loss in every nonzero contrast direction; the projection $P$ appears because a common displacement of all coordinates changes no ordering.

\subsection{The word versus the full configuration}\label{sec:full}

How much of the surviving information is lost by keeping only the word? The compressed coordinates answer this as well.

\begin{proposition}[Information in the full configuration]\label{prop:full}
In the setting of Theorem~\ref{thm:main},
\begin{equation}\label{eq:full}
I(U;r(t))=\frac1{2t}\,\E\bigl\|m_U-\bar m\bigr\|^2+O_{n,\lambda}(t^{-3/2}).
\end{equation}
\end{proposition}

\begin{proof}
Since $T$ is a bijection, $I(U;r(t))=I(U;M+N)$, with $N$ as in Lemma~\ref{lem:likelihood} and independent of $r(0)$. As in the proof of Theorem~\ref{thm:main}, change the origin and transport $f$ with it so that $M$ has nonnegative coordinates. Conditionally on $U=u$, the law of $M+N$ has density $L_u=\E[L_M\mid U=u]$ with respect to $Q_t$, and the unconditional density is $\bar L=\E L_M$. Write $\chi(w)=w\log w-w+1\ge0$. Because $\E_{Q_t}L_u=\E_{Q_t}\bar L=1$,
\[
I(U;M+N)=\sum_u\omega_u\,\E_{Q_t}\bigl[\bar L\,\chi(L_u/\bar L)\bigr].
\]
The integrand is defined as zero where $\bar L=0$, since every $L_u$ of positive prior probability then also vanishes.
By Lemma~\ref{lem:likelihood}, $\Lambda_u:=L_u-\bar L=t^{-1/2}(m_u-\bar m)\cdot V+\operatorname{rem}_{u,t}$ with $\|\operatorname{rem}_{u,t}\|_{L^p(Q_t)}=O(t^{-1})$ for every $p\in[1,\infty)$. Hence $\E_{Q_t}\Lambda_u^2=t^{-1}\|m_u-\bar m\|^2+O(t^{-3/2})$ and $\E_{Q_t}|\Lambda_u|^3=O(t^{-3/2})$. Let $\mathcal G_t=\{\max_i|V_i|\le t^{1/6}\}$. On $\mathcal G_t$ every factor in \eqref{eq:L} is $1+O(t^{-1/3})$, so $\bar L=1+O(t^{-1/3})$ uniformly, while $Q_t(\mathcal G_t^{\,c})=O(t^{-5})$ by Markov's inequality applied to a high moment of $V_i$. There is an absolute constant $c_0$ such that $|\chi(w)-\frac12(w-1)^2|\le c_0|w-1|^3$ for all $w\ge0$. Therefore, on $\mathcal G_t$ and for large $t$,
\[
\bar L\,\chi(L_u/\bar L)=\frac{\Lambda_u^2}{2\bar L}+O\bigl(|\Lambda_u|^3\bigr)=\tfrac12\Lambda_u^2\bigl(1+O(t^{-1/3})\bigr)+O\bigl(|\Lambda_u|^3\bigr).
\]
On $\mathcal G_t$ we also have $\bar L\ge1/2$ for large $t$. The expected error in replacing $\Lambda_u^2/(2\bar L)$ by $\Lambda_u^2/2$ there is at most $\|\bar L-1\|_{L^2(Q_t)}\|\Lambda_u\|_{L^4(Q_t)}^2=O(t^{-3/2})$ by Lemma~\ref{lem:likelihood}.
Off $\mathcal G_t$ we use $0\le\bar L\,\chi(L_u/\bar L)\le L_u\log(1/\omega_u)+\bar L$, which follows from $L_u\le\bar L/\omega_u$; by Cauchy--Schwarz the contribution of $\mathcal G_t^{\,c}$ to the expectation, and to $\E_{Q_t}\Lambda_u^2$, is $O(t^{-5/2})$. Collecting terms gives $\E_{Q_t}[\bar L\chi(L_u/\bar L)]=\frac1{2t}\|m_u-\bar m\|^2+O_{n,\lambda}(t^{-3/2})$, and summing over $u$ proves \eqref{eq:full}.
\end{proof}

\begin{corollary}[Ordering efficiency]\label{cor:efficiency}
If $\E\|m_U-\bar m\|^2>0$, then
\[
\lim_{t\to\infty}\frac{I(U;W_t)}{I(U;r(t))}=\frac{A_n}{n-1}\cdot\frac{\E\|P(m_U-\bar m)\|^2}{\E\|m_U-\bar m\|^2}\ \le\ \frac{A_n}{n-1}<1 .
\]
For fixed consecutive initial sites, a uniform initial word, and $U=W_0$, the limiting fraction is $A_n/n$.
\end{corollary}

\begin{proof}
Divide \eqref{eq:main} by \eqref{eq:full}. The inequalities follow from $\|Pv\|\le\|v\|$ and Lemma~\ref{lem:An}. For this choice $U=W_0$ we have $m_U=M$. In the consecutive uniform case $\Cov(M)$ is diagonal by the proof of Corollary~\ref{cor:Cn}, so $\tr(P\Cov(M))=(1-\frac1n)\tr\Cov(M)$.
\end{proof}

\paragraph{Degenerate coefficients.} When $C_U=0$, it is possible that $I(U;W_t)=0$ for every $t$ while
$I(U;r(t))$ decays as $t^{-1}$. For example, let $n=2$, let $U$ be uniform on $\{0,1\}$, and take $r(0)=(0,1)+U(1,1)$. Translation covariance gives $I(U;W_t)=0$ for every $t$, whereas \eqref{eq:full} gives $I(U;r(t))\sim1/(4t)$.

For consecutive sites and a uniform prior, $I(W_0;r(t))\sim n(n-1)(2n+5)/(144\,t)$, and the word alone retains the fraction $A_n/n$ of it: $1/\pi\approx0.318$ for $n=2$, $3/(2\pi)\approx0.477$ for $n=3$, and $0.79$ for $n=10$ (Table~\ref{tab:Cn}). By Lemma~\ref{lem:An} this fraction tends to one as $n\to\infty$, where the limit $t\to\infty$ is taken first at each fixed $n$.

\section{Exact two-particle channels and comparison with TASEP}\label{sec:two}

Let two distinct species start at distance $d\ge1$; $d$ is the distance between occupied sites, not the number of vacancies. Let $\Gamma_t$ be their distance and fix the initial ordering. Define
\[
h_k(t)=e^{-2t}\mathrm I_{|k|}(2t),\qquad k\in\Z,
\]
where $\mathrm I_j$ is the modified Bessel function of the first kind. Equivalently, $h_k(t)$ is the probability that the difference of two independent $\Pois(t)$ variables equals $k$.

\begin{proposition}[Long-range two-particle kernel]\label{prop:two}
For $k\ge1$,
\begin{align}
\PP(\Gamma_t=k,\,W_t=W_0)&=h_{k-d}(t),\label{eq:same}\\
\PP(\Gamma_t=k,\,W_t\ne W_0)&=h_{k+d-1}(t).\label{eq:rev}
\end{align}
In particular, the probability of reversed order is
\begin{equation}\label{eq:pd}
p_d(t)=\sum_{k=d}^\infty h_k(t)=\frac{1-h_0(t)-2\sum_{k=1}^{d-1}h_k(t)}{2}.
\end{equation}
\end{proposition}

\begin{proof}
Identify the state with the integer $\Gamma_t-1$ when the ordering agrees with the initial ordering and with the integer $-\Gamma_t$ when it is reversed. Away from adjacent particles, one clock increases the gap and the other decreases it, each at rate one. At distance one, a ring of the left particle's clock reverses the order
and leaves the distance equal to one. A ring of the right particle's
clock increases the distance to two. Thus the identified process is a nearest-neighbor walk on $\Z$, with rate one in each direction and initial state $d-1$. Its transition kernel is $h$. This proves \eqref{eq:same}--\eqref{eq:rev}; summation and symmetry give \eqref{eq:pd}.
\end{proof}

With the weaker label denoted by $a$ and the stronger by $b$, the ordering channel in the basis $(ab,ba)$ is
\begin{equation}\label{eq:KLR}
K_t^{\LR}=\begin{pmatrix}1-p_d(t)&p_d(t)\\ p_d(t)&1-p_d(t)\end{pmatrix}.
\end{equation}
The probability $p_d(t)$ records the order at time $t$; it is not the probability of at least one exchange.

For comparison, in ordinary multispecies TASEP with rate-one clocks a particle moves one site to the right into a vacancy, and a nearest-neighbor priority exchange $ba\to ab$ is permitted. The order $ab$ can never become $ba$. Starting in $ba$, the gap before the first exchange is a symmetric walk on the positive integers killed at zero. Reflection gives its killed kernel $h_{k-d}(t)-h_{k+d}(t)$ at final distance $k$. Its survival probability is therefore
\begin{equation}\label{eq:sd}
s_d(t)=h_0(t)+2\sum_{k=1}^{d-1}h_k(t)+h_d(t),\qquad K_t^{\ord}=\begin{pmatrix}1&0\\1-s_d(t)&s_d(t)\end{pmatrix}.
\end{equation}

\begin{theorem}[Different information exponents]\label{thm:two}
For a uniform initial ordering at fixed occupied sites,
\begin{align}
I_{\LR}(W_0;W_t)&=\log2-\hb(p_d(t))\sim\frac{(2d-1)^2}{8\pi t},\label{eq:ILR}\\
I_{\ord}(W_0;W_t)&=\hb(s_d(t)/2)-\tfrac12\hb(s_d(t))\sim\frac{d\log2}{2\sqrt\pi}\,t^{-1/2},\label{eq:Iord}
\end{align}
where $\hb(u)=-u\log u-(1-u)\log(1-u)$.
\end{theorem}

\begin{proof}
The exact information expressions follow from the two matrices. For fixed integer $k$, the Bessel asymptotic gives
\[
h_k(t)=\frac1{\sqrt{4\pi t}}\bigl(1+O_k(t^{-1})\bigr).
\]
Hence
\[
1-2p_d(t)\sim\frac{2d-1}{2\sqrt{\pi t}},\qquad s_d(t)\sim\frac d{\sqrt{\pi t}}.
\]
Writing $\gamma=1-2p_d(t)$, the expansion $\log2-\hb((1-\gamma)/2)=\gamma^2/2+O(\gamma^4)$ proves \eqref{eq:ILR}. Also,
\[
\hb(s/2)-\tfrac12\hb(s)=\frac{\log2}2s+O(s^2),\qquad s\downarrow0,
\]
which proves \eqref{eq:Iord}.
\end{proof}

\begin{figure}[H]
\centering
\begin{tikzpicture}
\begin{loglogaxis}[width=0.74\textwidth,height=6.2cm,
  xlabel={Time $t$},ylabel={$I(W_0;W_t)$ (nats)},
  xmin=0.03,xmax=1200,ymin=2e-5,ymax=1.2,
  legend pos=south west,legend cell align=left,legend style={font=\footnotesize,draw=black!30},
  tick label style={font=\footnotesize},label style={font=\small},
  grid=major,grid style={black!10}]
\addplot[lrblue,very thick] coordinates {(0.031623,0.5578) (0.038178,0.537862) (0.046093,0.515588) (0.055649,0.490888) (0.067185,0.463734) (0.081113,0.434176) (0.097928,0.402372) (0.11823,0.368604) (0.14274,0.333301) (0.17233,0.297046) (0.20806,0.260566) (0.25119,0.224705) (0.30326,0.190364) (0.36613,0.158422) (0.44203,0.129632) (0.53367,0.104529) (0.6443,0.083356) (0.77787,0.0660478) (0.93913,0.0522688) (1.1338,0.0415029) (1.3689,0.0331647) (1.6527,0.0266978) (1.9953,0.021637) (2.4089,0.0176286) (2.9083,0.0144182) (3.5112,0.0118251) (4.2391,0.00971798) (5.1179,0.00799868) (6.1789,0.00659148) (7.4598,0.00543703) (9.0063,0.00448819) (10.873,0.00370721) (13.127,0.00306365) (15.849,0.00253284) (19.135,0.00209469) (23.101,0.00173281) (27.89,0.00143376) (33.672,0.00118654) (40.653,0.000982099) (49.081,0.000812981) (59.255,0.000673055) (71.539,0.000557259) (86.37,0.000461417) (104.28,0.000382081) (125.89,0.000316401) (151.99,0.000262022) (183.5,0.000216996) (221.54,0.000179712) (267.47,0.000148837) (322.92,0.000123269) (389.86,0.000102095) (470.68,8.45591e-05) (568.26,7.00358e-05) (686.06,5.80075e-05) (828.29,4.80453e-05) (1000,3.97942e-05)};
\addlegendentry{Long-range swap}
\addplot[ordred,very thick] coordinates {(0.031623,0.624183) (0.038178,0.613708) (0.046093,0.601876) (0.055649,0.588578) (0.067185,0.573713) (0.081113,0.557201) (0.097928,0.538984) (0.11823,0.519044) (0.14274,0.497406) (0.17233,0.474152) (0.20806,0.449428) (0.25119,0.42345) (0.30326,0.396506) (0.36613,0.368952) (0.44203,0.341194) (0.53367,0.313674) (0.6443,0.286833) (0.77787,0.26108) (0.93913,0.236758) (1.1338,0.214116) (1.3689,0.193294) (1.6527,0.174328) (1.9953,0.157167) (2.4089,0.141704) (2.9083,0.127803) (3.5112,0.115317) (4.2391,0.104104) (5.1179,0.0940314) (6.1789,0.0849788) (7.4598,0.0768376) (9.0063,0.0695111) (10.873,0.0629129) (13.127,0.0569663) (15.849,0.0516032) (19.135,0.0467631) (23.101,0.0423922) (27.89,0.0384425) (33.672,0.0348716) (40.653,0.0316412) (49.081,0.0287175) (59.255,0.0260702) (71.539,0.0236721) (86.37,0.0214989) (104.28,0.0195288) (125.89,0.0177422) (151.99,0.0161215) (183.5,0.0146509) (221.54,0.0133162) (267.47,0.0121045) (322.92,0.0110042) (389.86,0.0100049) (470.68,0.00909717) (568.26,0.00827247) (686.06,0.00752309) (828.29,0.00684206) (1000,0.00622306)};
\addlegendentry{Ordinary mTASEP}
\addplot[lrblue,dashed,domain=2:1000,samples=2] {1/(8*pi*x)};
\addlegendentry{$1/(8\pi t)$}
\addplot[ordred,dashed,domain=2:1000,samples=30] {ln(2)/(2*sqrt(pi*x))};
\addlegendentry{$(\log 2)/(2\sqrt{\pi t})$}
\end{loglogaxis}
\end{tikzpicture}
\caption{Exact two-particle information for adjacent initial sites and a uniform initial ordering. Solid curves are the formulas in Theorem~\ref{thm:two}; dashed curves are the proved leading asymptotics.}
\label{fig:two}
\end{figure}
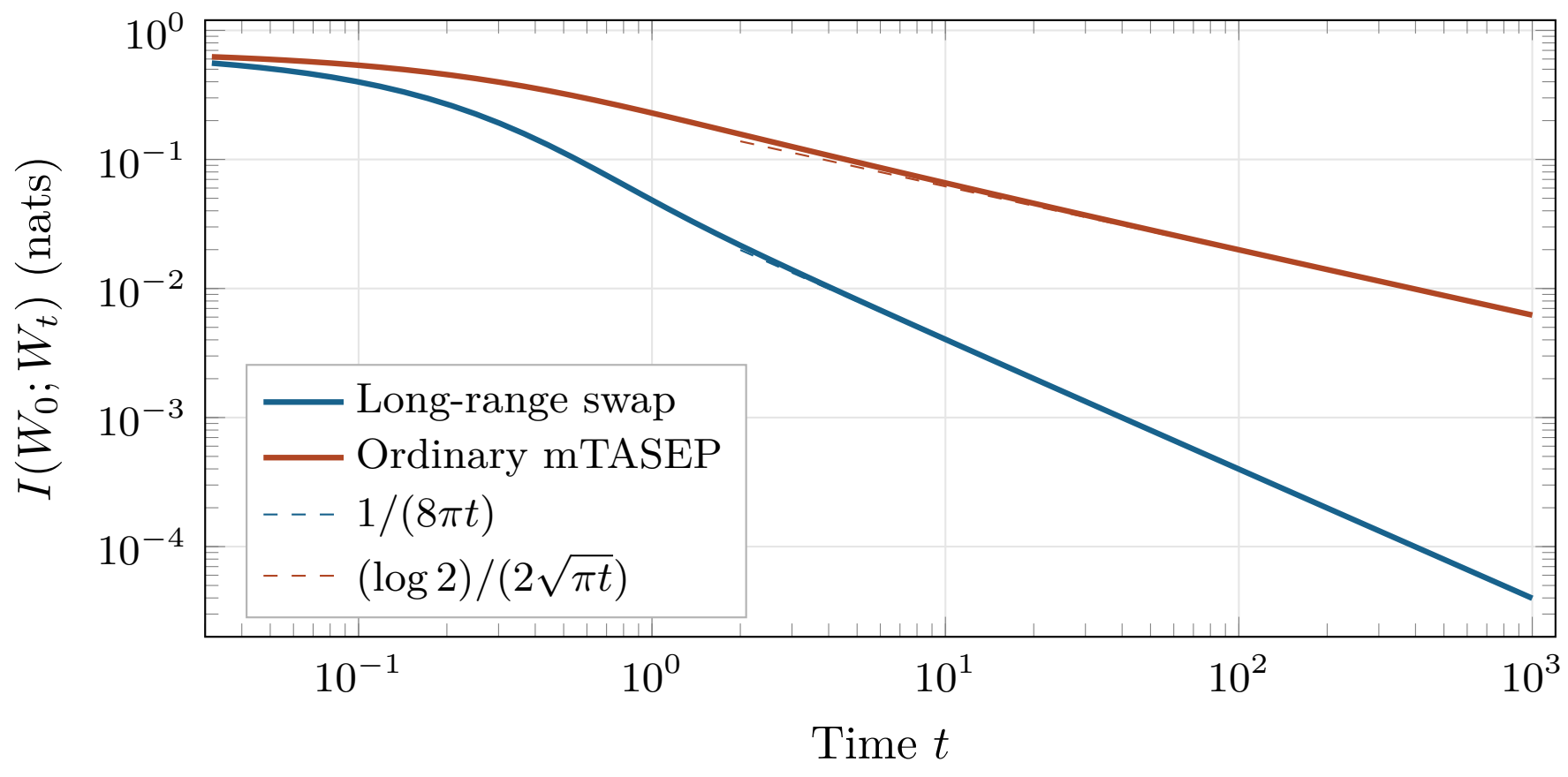

Figure~\ref{fig:two} shows both exact expressions for adjacent initial sites, $d=1$, together with their leading asymptotics.

\paragraph{A gap varying with time.}\label{par:gapscale}
For a uniform initial word, \eqref{eq:ILR} holds in the quantitative form
\begin{equation}\label{eq:gaprate}
 I_{\LR}(W_0;W_t)=\frac{(2d-1)^2}{8\pi t}\{1+O(d^2/t)\}
\end{equation}
whenever $t\to\infty$ and the integer gap satisfies $d=o(\sqrt t)$. Indeed, the Fourier
representation $h_k(t)=\frac1{2\pi}\int_{-\pi}^{\pi}e^{-2t(1-\cos u)}\cos(ku)\,du$ and the bound
$1-\cos u\ge2u^2/\pi^2$ on $[-\pi,\pi]$ give $0\le h_0(t)-h_k(t)\le Ck^2t^{-3/2}$. Summing over
$1\le k<d$ in \eqref{eq:pd} and using $h_0(t)=(4\pi t)^{-1/2}(1+O(t^{-1}))$ gives
$1-2p_d=(2d-1)(4\pi t)^{-1/2}(1+O(d^2/t))$, and the binary entropy expansion proves
\eqref{eq:gaprate}. When $d/\sqrt t\to c>0$ the expansion fails:
$1-2p_d=\PP(|N_1-N_2|\le d-1)$ then tends to $2\Phi(c/\sqrt2)-1$, for independent $\Pois(t)$
variables $N_1,N_2$.

\paragraph{A nonzero signal with vanishing first coefficient.}\label{par:degeneratetwo}
Let the two particles start at $(0,d)$ in species order $12$. An equiprobable binary input
selects the gaps $d=1$ and $d=5$ with equal conditional probabilities for input zero, and the
gap $d=3$ for input one. Both conditional mean compressed vectors equal $(0,3)$, so $C_U=0$.
The large-argument expansion of the modified Bessel function gives
\[
 h_k(t)=\frac1{2\sqrt{\pi t}}\left(1-\frac{4k^2-1}{16t}
                  +O_k(t^{-2})\right),
\]
so by \eqref{eq:pd} the coefficient of $t^{-3/2}$ in $p_d$ is
$[-1+2\sum_{k=1}^{d-1}(4k^2-1)]/(64\sqrt\pi)$, with numerators $-1,35,231$ for $d=1,3,5$.
The constant and $t^{-1/2}$ terms cancel in $\Delta=(p_1(t)+p_5(t))/2-p_3(t)$, leaving
$\Delta=5t^{-3/2}/(4\sqrt\pi)+O(t^{-5/2})$. Both rows converge to $(1/2,1/2)$, and about their
mean the two reversal probabilities deviate by $\pm\Delta/2$. Hence
\[
 I(U;W_t)\sim\frac{\Delta^2}{2}\sim\frac{25}{32\pi t^3}.
\]
Since $W_0=12$ for both inputs, $I(U;W_0)=0$ here. Positive later
information is compatible with Theorem~\ref{thm:main}.

We now fix the initial gap $d$ and take the initial word $W_0$ as the input. The total variation distance between the two rows is $1-2p_d(t)\sim(2d-1)/(2\sqrt{\pi t})$ in the long-range model and $s_d(t)\sim d/\sqrt{\pi t}$ in the ordinary model, so the two models lose distinguishability with the same algebraic exponent. For every prior on the two initial words, $I(W_0;W_t)$ is nonincreasing in $t$ in both models.

\begin{proposition}[Monotonicity for two particles]\label{prop:tv2}
Consider two particles at distance $d$. The function $p_d$ is increasing with $p_d<\frac12$, and $s_d$ is nonincreasing. For $0\le t\le t'$ the channel at time $t'$ is, in both models, the channel at time $t$ followed by a further channel. Consequently $I(W_0;W_t)$ is nonincreasing in $t$ for every prior.
\end{proposition}

\begin{proof}
By the proof of Proposition~\ref{prop:two}, $p_d(t)$ is the probability that a symmetric walk started at $d-1$ is at or below $-1$ at time $t$. Its derivative is the net probability flux across the bond between $0$ and $-1$, namely $h_{d-1}(t)-h_d(t)$, which is positive for $t>0$ because $\mathrm I_k(x)$ is decreasing in $k\ge0$ for $x>0$. Also $p_d<\frac12$ by \eqref{eq:pd}, and $s_d$ is a survival probability. For $p\le p'<\frac12$ the binary symmetric channel with crossover probability $p'$ is the one with crossover $p$ followed by the one with crossover $(p'-p)/(1-2p)$. For $0<s\leq 1$  and $0\leq s'\le s$, the matrix $K^{\ord}$ with parameter $s'$ is the product of the matrices with parameters $s$ and $s'/s$. The data-processing inequality gives the last claim.
\end{proof}

The different exponents have a simple channel interpretation. The long-range output converges to the interior distribution $(1/2,1/2)$, so the leading relative entropy is quadratic in the row difference. For the two distinct species the ordinary output converges to the boundary point $(1,0)$; information about the initial word survives in a rare unsorted output, giving a term linear in its probability. For $\PP(W_0=ba)=\theta\in(0,1)$, the ordinary channel in \eqref{eq:sd}
gives $I_{\ord}(W_0;W_t)\sim-\theta\log(\theta)\,d/\sqrt{\pi t}$ by expanding
$h_{\rm b}(\theta s_d)-\theta h_{\rm b}(s_d)$ at $s_d=0$.

\section{Species composition, diversity and information loss}\label{sec:types}

Repeated species distinguish the two rules. Throughout this section the species counts
are fixed and all clocks have rate one.

\subsection{Composition entropy and initial ordering information}

Fix positive multiplicities $\boldsymbol n=(n_1,\ldots,n_q)$, listed from weak to strong, with $\sum_i n_i=n$, and let $\mathcal W_{\boldsymbol n}$ be the set of words with these multiplicities. Two entropies are attached to $\boldsymbol n$:
\begin{equation}\label{eq:compositionentropies}
H_{\rm comp}(\boldsymbol n)=-\sum_i\frac{n_i}{n}\log\frac{n_i}{n},\qquad
S_{\rm ord}(\boldsymbol n)=\log\frac{n!}{\prod_i n_i!}.
\end{equation}
The first is the entropy of the species of a uniformly chosen particle; it measures composition diversity and is conserved by the dynamics. The second is the logarithm of the number of words, so $H(W_0)=S_{\rm ord}$ under the uniform word prior. Both vanish for a single species and take their maxima $\log n$ and $\log n!$ for distinct species. When the particles start on consecutive sites with a uniform word and every species follows the same rule, we write $I_{\boldsymbol n}^{\alpha}(t)=I(W_0;W_t)$, with $\alpha=0$ for the TASEP-type rule and $\alpha=1$ for the drop--push-type rule.

\subsection{Reduction to independent single-species processes}

Let species $i$ follow the rule $\alpha_i\in\{0,1\}$ and write $x_{i,1}<\cdots<x_{i,n_i}$ for its positions. As in \eqref{eq:T}, delete the sites of stronger species:
\begin{equation}\label{eq:blockcoords}
z_{i,k}=x_{i,k}-\#\{(j,\ell):j>i,\ x_{j,\ell}<x_{i,k}\},
\qquad 1\le k\le n_i.
\end{equation}
These coordinates are strictly increasing in $k$. On $\mathcal Z_m=\{z\in\Z^m:z_1<\cdots<z_m\}$ we use two rate-one single-species processes. In TASEP an activated particle moves one site to the right if that site is empty. In the drop--push process it moves to the first empty site on its right; equivalently, the run of consecutive occupied sites starting at its position shifts one step to the right. When these processes are applied to the compressed coordinates of species $i$, a site is empty if it holds no particle of species $i$; it may hold a weaker particle.

\begin{proposition}[Species-wise coordinate reduction]\label{prop:blockreduction}
The map \eqref{eq:blockcoords} is a bijection from the configurations with multiplicities $\boldsymbol n$ onto $\prod_i\mathcal Z_{n_i}$. An activation of a particle of species $i$ performs the single-species move of that particle in $(z_{i,k})_k$, of TASEP type if $\alpha_i=0$ and of drop--push type if $\alpha_i=1$, and leaves the coordinates of every other species unchanged. Consequently, given the initial configuration, the compressed species evolve as independent single-species processes.
\end{proposition}

\begin{proof}
To invert the map, insert the species in decreasing order. If $S$ is the set of sites occupied by the species stronger than $i$, then $x\mapsto x-\#\{h\in S:h<x\}$ is an increasing bijection from $\Z\setminus S$ onto $\Z$, and the particles of species $i$ go to the preimages of their $z_{i,k}$.

Let a particle of species $i$ ring at $x$ with target $y$. The labels strictly between $x$ and $y$ are greater than $i$ under the TASEP-type rule and at least $i$ under the drop--push-type rule. After the stronger labels are deleted, the particle therefore steps to the next site if it is empty, stays put if that site holds an equal particle and $\alpha_i=0$, and jumps to the first empty site if $\alpha_i=1$. If the target holds a weaker particle, that particle moves from $y$ to $x$ and loses the $y-x$ stronger particles of $[x,y)$ from its left, so its coordinate is unchanged. Every other particle keeps its site and its number of stronger particles on the left: a particle inside $(x,y)$ is at least as strong as both moving labels, and a particle outside $[x,y]$ sees them on the same side as before. All rates are one, so the generator is the sum of the single-species generators. The rates are bounded, so the process is the product of independent single-species chains.
\end{proof}

For distinct species the factors are the Poisson walkers of Theorem~\ref{thm:poisson}. Only Theorem~\ref{thm:repeated} uses this reduction; the short-time results rely on the scan rule alone.

\subsection{Initial information loss for every composition}

\begin{theorem}[Composition-dependent initial loss]\label{thm:composition}
Start from the consecutive sites $0,\ldots,n-1$ with $W_0$ uniform on $\mathcal W_{\boldsymbol n}$, and let species $i$ follow the rule $\alpha_i\in\{0,1\}$. Fix an initial word $w$. For each occupied site $x$, consider the species
word obtained if the clock of the particle at $x$ rings first. For
$v\ne w$, let $k(w,v)$ be the number of sites $x$ for which this word
is $v$, and put $\kappa(w)=\sum_{v\ne w}k(w,v)$. As $t\downarrow0$,
\begin{equation}\label{eq:compositionexpansion}
I(W_0;W_t)=S_{\rm ord}(\boldsymbol n)-r\,t\log(1/t)-c\,t+O_{\boldsymbol n}(t^2\log(1/t)),
\end{equation}
where, with $0\log0=0$,
\begin{equation}\label{eq:compositionlinear}
r=\E\,\kappa(W_0),\qquad
c=r-\E\sum_{v\ne W_0}k(W_0,v)\log k(W_0,v).
\end{equation}
With $n_{\le i}=n_1+\cdots+n_i$,
\begin{equation}\label{eq:mixedrates}
 r=n-\sum_{i:\alpha_i=0}\left(\frac{n_i}{n}+\frac{n_i(n_i-1)}{n_{\le i}}\right)
 -\sum_{i:\alpha_i=1}\frac{n_i}{n-n_i+1}.
\end{equation}
In particular, if every species follows the TASEP-type rule, or every species the drop--push-type rule, then $r$ equals respectively
\begin{equation}\label{eq:compositionrates}
r_0(\boldsymbol n)=n-1-\sum_i\frac{n_i(n_i-1)}{n_{\le i}},\qquad
r_1(\boldsymbol n)=n-\sum_i\frac{n_i}{n-n_i+1}.
\end{equation}
\end{theorem}

\begin{proof}
Write $K_t(w,v)=\PP(W_t=v\mid W_0=w)$ for $w,v\in\mathcal W_{\boldsymbol n}$ and $\mu_t$ for the law of $W_t$. By uniformization at total rate $n$, $K_t(w,v)=k(w,v)t+O(t^2)$ for $v\ne w$, and $K_t(w,w)=1-\kappa(w)t+O(t^2)$. In the row entropy an entry $kt+O(t^2)$ with $k\ge1$ contributes $kt\log(1/t)-kt\log k+O(t^2\log(1/t))$, an entry $O(t^2)$ contributes $O(t^2\log(1/t))$, and the diagonal contributes $\kappa(w)t+O(t^2)$. Hence
\[
H(K_t(w,\cdot))=\kappa(w)\,t\log(1/t)
+\Bigl[\kappa(w)-\sum_{v\ne w}k(w,v)\log k(w,v)\Bigr]t
+O(t^2\log(1/t)).
\]
The output law $\mu_t$ differs from the uniform law $\mathsf U$ on $\mathcal W_{\boldsymbol n}$ by $O(t)$, and relative entropy is quadratic in such a perturbation, so $S_{\rm ord}-H(W_t)=D(\mu_t\|\mathsf U)=O(t^2)$. Subtracting the average row entropy from $H(W_t)$ proves \eqref{eq:compositionexpansion} and \eqref{eq:compositionlinear}.

It remains to count the activations that leave a packed word unchanged. Under the TASEP-type rule a particle of species $i$ leaves the word unchanged exactly when it is the rightmost particle, or the nearest particle on its right with label at most $i$ is again an $i$. The rightmost particle has species $i$ with probability $n_i/n$. For the second case, delete the labels greater than $i$: the remaining word is uniform with multiplicities $(n_1,\ldots,n_i)$, and each of its $n_{\le i}-1$ bonds, if any, is a pair $ii$ with probability $n_i(n_i-1)/(n_{\le i}(n_{\le i}-1))$. The expected number of unchanged clocks of species $i$ is thus $n_i/n+n_i(n_i-1)/n_{\le i}$. Under the drop--push-type rule a particle of species $i$ leaves the word unchanged exactly when every particle on its right is also an $i$. The $n_i$ copies fall into the $n-n_i+1$ gaps around the other particles with exchangeable counts, so the expected number in the last gap is $n_i/(n-n_i+1)$. Each activation is governed by the rule of its own species, which proves \eqref{eq:mixedrates}.
\end{proof}

\subsection{Diversity, priority and the choice of rule}

\begin{corollary}[Refinement and priority placement]\label{cor:diversity}
Let every species follow the same rule, with $r_0,r_1$ as in \eqref{eq:compositionrates}.
\begin{enumerate}
\item [(i)] $r_1\ge r_0$, with strict inequality if there are at least two species and one of them is repeated. In that case $I_{\boldsymbol n}^{1}(t)<I_{\boldsymbol n}^{0}(t)$ for all sufficiently small $t>0$.
\item [(ii)] Splitting a species of size $a+b$ into a weaker species of size $a$ and a stronger one of size $b$ at consecutive priority levels increases $H_{\rm comp}$, $S_{\rm ord}$, $r_0$ and $r_1$.
\item [(iii)] The coefficient $r_1$ depends only on the multiset of multiplicities. Among its orderings, $r_0$ is largest when the multiplicities are nondecreasing from weak to strong and smallest when they are nonincreasing.
\end{enumerate}
\end{corollary}

\begin{proof}
(i) A word-changing TASEP-type activation meets no equal label before it stops, so it acts in the same way under the drop--push-type rule. Hence $r_1\ge r_0$. Under the stated conditions some packed word contains two adjacent copies $ii$ followed by a different label, and the first copy changes the word only under the drop--push-type rule. The inequality $I_{\boldsymbol n}^{1}(t)<I_{\boldsymbol n}^{0}(t)$ follows from \eqref{eq:compositionexpansion}.

(ii) The entropy increments are $\Delta H_{\rm comp}=[(a+b)\log(a+b)-a\log a-b\log b]/n>0$ and $\Delta S_{\rm ord}=\log\binom{a+b}{a}>0$. If $s$ particles are weaker than both new species, \eqref{eq:compositionrates} gives
\[
\Delta r_0=\frac{ab(2s+a+1)}{(s+a)(s+a+b)}>0 .
\]
With $f(x)=x/(n+1-x)$ we have $\Delta r_1=f(a+b)-f(a)-f(b)>0$, since $f(x)/x$ is increasing.

(iii) The formula for $r_1$ is symmetric in the $n_i$. If adjacent multiplicities $a,b$ are preceded by $s$ weaker particles, then
\[
r_0(\ldots,a,b,\ldots)-r_0(\ldots,b,a,\ldots)
=\frac{ab(b-a)(s+1)}{(s+a)(s+b)(s+a+b)},
\]
and sorting by adjacent swaps gives both extrema.
\end{proof}

Composition entropy alone therefore does not determine the information loss. The compositions $(2,1)$ and $(1,2)$ have the same two entropies, but $r_0=1$ and $4/3$, respectively. Under the drop--push-type rule both have $r_1=5/3$, and they differ at the next order: \eqref{eq:compositionlinear} gives $c=(5-2\log2)/3$ and $c=5/3$, because two activations of the word $112$ produce $121$. For four particles, the compositions $(2,1,1)$, $(1,2,1)$ and $(1,1,2)$ have $r_0=2$, $7/3$ and $5/2$, and the same $r_1=17/6$.

Normalizing by the initial information can reverse a comparison. Refining $(2,1)$ or $(1,2)$ gives $(1,1,1)$, with $r_0=2$ and $S_{\rm ord}=\log6$. Since $1/\log3<2/\log6<4/(3\log3)$, the fractional loss $r_0/S_{\rm ord}$ increases in the first case and decreases in the second, although both absolute losses increase. Among the 741 splits with $n\le8$ and at least two species before splitting, an exact enumeration (Supplement~S1) shows that $r_0/S_{\rm ord}$ increases in 358 and decreases in 383, whereas $r_1/S_{\rm ord}$ decreases in every case.

\subsection{Long-time information for a repeated pair}
\label{sec:repeatasym}

Take three particles on the sites $0,1,2$ with a uniform initial word, one species occurring twice. Let $\mathsf L_t$ be the number of particles to the left of the singleton at time $t$. Its values $0,1,2$ label the three words, so $I(W_0;W_t)=I(\mathsf L_0;\mathsf L_t)$. For the repeated pair we call the TASEP-type rule \emph{blocking} and the drop--push-type rule \emph{pushing}.

\begin{theorem}[One repeated species]\label{thm:repeated}
Let $\boldsymbol n=(2,1)$ or $(1,2)$, with counts listed from weak to strong.
For each initial word, the probabilities of finding the singleton in the left,
middle and right positions converge under the TASEP-type rule to
\[
 (p_{\rm L},p_{\rm M},1/2)
 =\left(\frac13-\frac{\sqrt3}{4\pi},\,
         \frac16+\frac{\sqrt3}{4\pi},\,\frac12\right),
\]
and under the drop--push-type rule to the same vector in reverse order, in both
cases with error $O(t^{-1/2})$. With
$\mathcal J=(9/p_{\rm L}+1/p_{\rm M}+32)/(64\pi)$, both rules satisfy
\begin{equation}\label{eq:repeatedcoef}
 t I_{(2,1)}^\alpha(t)=\frac{\mathcal J}{3}+O(t^{-1/2}),\qquad
 t I_{(1,2)}^\alpha(t)=\frac{\mathcal J}{9}+O(t^{-1/2})
 \qquad(t\to\infty).
\end{equation}
\end{theorem}

\begin{figure}[H]
\centering
\pgfplotstableread[col sep=space,row sep=\\]{
% BEGIN DISTINCT DATA
t three four\\
0.125 0.13515331345 0.26033769773\\
0.25 0.18968938998 0.38738385011\\
0.5 0.22274187619 0.49485772552\\
1 0.22895827024 0.5599843999\\
2 0.22651207035 0.59835169498\\
4 0.22388137086 0.61610790273\\
8 0.22217465983 0.62274460522\\
16 0.22107881327 0.62479310384\\
64 0.21988458593 0.6246178408\\
256 0.21934220009 0.62350679691\\
1024 0.21908524021 0.62270461994\\
% END DISTINCT DATA
}\distinctdata
\pgfplotstableread[col sep=space,row sep=\\]{
% BEGIN REPEATED DATA
t weakBlocked weakPush strongBlocked strongPush\\
0.125 0.09131766826 0.073806076099 0.077974346833 0.067683696357\\
0.25 0.13639319309 0.09977319422 0.10188495681 0.080927430421\\
0.5 0.17068187629 0.11686422971 0.10051571774 0.073012449233\\
1 0.17664226658 0.12455499151 0.074452164011 0.054984793834\\
2 0.16663826814 0.1276796858 0.055280505997 0.045112549566\\
4 0.15673031623 0.12940872807 0.048560043801 0.041763895223\\
8 0.14981827885 0.13073369146 0.046062224012 0.041039453894\\
16 0.14515356223 0.13178271526 0.045076965764 0.041341001571\\
64 0.13981553413 0.13320396507 0.04463061865 0.042621282338\\
256 0.13727309455 0.13398528316 0.044692497551 0.043647002302\\
1024 0.13603415836 0.13439467433 0.044792758818 0.044259110664\\
4096 0.13542280846 0.13460416282 0.044860120786 0.044590480906\\
16384 0.13511916749 0.13471011765 0.044898106057 0.044762570465\\
% END REPEATED DATA
}\repeateddata
\begin{minipage}[t]{0.485\linewidth}\vspace{0pt}
\centering
\begin{tikzpicture}
\begin{semilogxaxis}[
 width=\linewidth,height=5.6cm,
 xlabel={Time $t$},ylabel={$tI(W_0;W_t)$},
 title={(a) Distinct species},title style={font=\small},
 tick label style={font=\scriptsize},label style={font=\small},
 xmin=0.09,xmax=1550,ymin=0,ymax=0.68,
 xtick={0.1,1,10,100,1000},xticklabels={$10^{-1}$,$1$,$10$,$10^2$,$10^3$},
 ytick={0,0.2,0.4,0.6},yticklabel style={/pgf/number format/fixed},
 ymajorgrids=true,grid style={black!12},mark size=1.3pt,
 legend style={draw=none,fill=none,font=\scriptsize,at={(0.5,-0.29)},anchor=north,
 legend columns=2,/tikz/every even column/.append style={column sep=0.4em}}]
\draw[lrblue,densely dashed] (axis cs:0.125,0.21883804675)--(axis cs:1024,0.21883804675);
\draw[ordred,densely dashed] (axis cs:0.125,0.62174093570)--(axis cs:1024,0.62174093570);
\addplot[lrblue,thick,mark=*] table[x=t,y=three]{\distinctdata};
\addlegendentry{$n=3$}
\addplot[ordred,thick,mark=square*] table[x=t,y=four]{\distinctdata};
\addlegendentry{$n=4$}
\end{semilogxaxis}
\end{tikzpicture}
\end{minipage}\hfill
\begin{minipage}[t]{0.485\linewidth}\vspace{0pt}
\centering
\begin{tikzpicture}
\begin{semilogxaxis}[
 width=\linewidth,height=5.6cm,
 xlabel={Time $t$},ylabel={$tI(W_0;W_t)$},
 title={(b) A repeated species},title style={font=\small},
 tick label style={font=\scriptsize},label style={font=\small},
 xmin=0.09,xmax=24000,ymin=0,ymax=0.19,
 xtick={0.1,1,10,100,1000,10000},xticklabels={$10^{-1}$,$1$,$10$,$10^2$,$10^3$,$10^4$},
 ytick={0,0.05,0.10,0.15},yticklabel style={/pgf/number format/fixed,/pgf/number format/precision=2},
 ymajorgrids=true,grid style={black!12},mark size=1.2pt,
 legend style={draw=none,fill=none,font=\scriptsize,at={(0.5,-0.29)},anchor=north,
 legend columns=2,/tikz/every even column/.append style={column sep=0.4em}}]
\draw[black!70,densely dotted,semithick] (axis cs:0.125,0.13481688396)--(axis cs:16384,0.13481688396);
\draw[black!70,densely dotted,semithick] (axis cs:0.125,0.044938961319)--(axis cs:16384,0.044938961319);
\addplot[lrblue,thick,mark=*] table[x=t,y=weakBlocked]{\repeateddata};
\addlegendentry{$(2,1)$ TASEP}
\addplot[lrblue,thick,dashed,mark=o] table[x=t,y=weakPush]{\repeateddata};
\addlegendentry{$(2,1)$ push}
\addplot[ordred,thick,mark=square*] table[x=t,y=strongBlocked]{\repeateddata};
\addlegendentry{$(1,2)$ TASEP}
\addplot[ordred,thick,dashed,mark=square] table[x=t,y=strongPush]{\repeateddata};
\addlegendentry{$(1,2)$ push}
\end{semilogxaxis}
\end{tikzpicture}
\end{minipage}
\caption{Scaled ordering information for consecutive sites and a uniform initial word.
In (a) the horizontal dashed lines are the proved limits $C_3$ and $C_4$.
In (b) the dotted lines are the two limits in \eqref{eq:repeatedcoef},
the counts are listed from weak to strong, and solid and dashed curves show
the TASEP-type and drop--push-type rules, respectively. Markers are numerical
evaluations of the exact finite-time formulas (Supplement~S1).}\label{fig:late}
\end{figure}
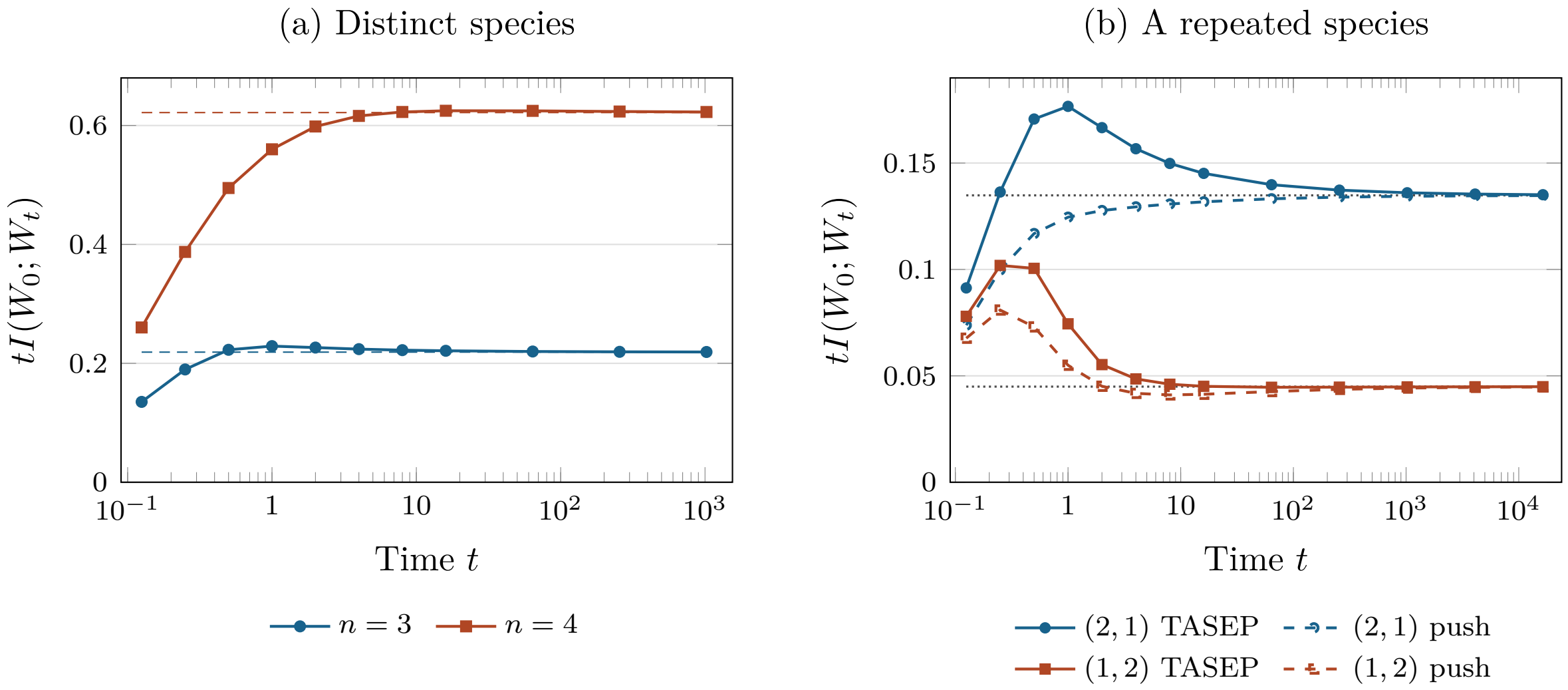

The two compositions have the same $H_{\rm comp}$ and the same $H(W_0)=\log3$, yet repeating the weaker species triples the coefficient. The proof shows where the factor comes from. To leading order the three initial words act as three starting sites of the singleton relative to a fixed pair. In the order $\mathsf L_0=0,1,2$ these are $0,1,2$ for $(2,1)$. For $(1,2)$ they are $0,0,1$ under blocking and $0,1,1$ under pushing, so two of the three words give the same row up to $O(t^{-1})$. The variances are $2/3$ and $2/9$. Numerically $\mathcal J/3\approx0.134817$ and $\mathcal J/9\approx0.044939$. At $t=16384$ the computed values of $tI$ are $0.13512$ and $0.13471$ for $(2,1)$, and $0.04490$ and $0.04476$ for $(1,2)$, under blocking and pushing respectively (Figure~\ref{fig:late}(b)).

The proof uses the marginal laws of a pair of equal particles. Write
$f_c(x)=\PP(c+N=x)$ and $F_c(x)=\PP(c+N\le x)$ for $N\sim\Pois(t)$ and integers $c,x$.

\begin{lemma}[Marginals of a pair]\label{lem:pair}
Let two particles of one species start at $a<b$ and let $X_-<X_+$ be their positions at time $t$ in the single-species process on $\mathcal Z_2$. Under blocking $X_+$ has distribution function $F_b$, and under pushing $X_-$ has distribution function $F_a$. The other two marginals are
\begin{equation}\label{eq:paircdfs}
\begin{aligned}
 \PP(X_-\le x)&=1-[1-F_a(x)][1-F_b(x)]+f_b(x+1)\,\E[(a+N-x)_+]&&\text{\upshape(blocking)},\\
 \PP(X_+\le x)&=F_a(x)F_b(x)-f_a(x)\,\E[(x+1-b-N)_+]&&\text{\upshape(pushing)}.
\end{aligned}
\end{equation}
\end{lemma}

\begin{proof}
Under blocking the front position, and under pushing the rear position, increases by one at every ring of the clock of the particle occupying it. This gives the two free marginals. From $(a,b)$ to $x<y$ the pair kernels are the two-particle cases of the single-species laws for TASEP \cite{Schutz-1997} and for the drop--push process \cite{Ali},
\[
\begin{aligned}
 Q_0(x,y)&=f_a(x)f_b(y)-[f_b(x)-f_b(x+1)][1-F_a(y-1)],\\
 Q_1(x,y)&=f_a(x)f_b(y)-F_b(x)[f_a(y)-f_a(y-1)].
\end{aligned}
\]
Indeed, every factor $u$ satisfies $\partial_tu(k)=u(k-1)-u(k)$, so both expressions solve the free forward equation. The identity $F_c(k)-F_c(k-1)=f_c(k)$ gives the blocking condition $Q_0(x,x)=Q_0(x,x+1)$ and the pushing condition $Q_1(x,x)=Q_1(x-1,x)$. At $t=0$ both corrections vanish on $x<y$, leaving the point mass at $(a,b)$. Both kernels are summable with their time derivatives, locally uniformly in $t$, so uniqueness for the forward equation of a bounded-rate chain identifies them. Summing over the free particle gives
\[
\begin{aligned}
 \PP(X_-=x)&=f_a(x)[1-F_b(x)]-[f_b(x)-f_b(x+1)]\,\E[(a+N-x)_+],\\
 \PP(X_+=y)&=f_b(y)F_a(y-1)-[f_a(y)-f_a(y-1)]\,\E[(y-b-N)_+].
\end{aligned}
\]
These are the first differences of the functions in \eqref{eq:paircdfs}, which vanish at $-\infty$.
\end{proof}

\begin{proof}[Proof of Theorem~\ref{thm:repeated}]
\emph{Rows.} Fix $j=\mathsf L_0$  and write $K_t(j,k)=\PP(\mathsf L_t=k\mid\mathsf L_0=j)$ for $k\in\{0,1,2\}$. By Proposition~\ref{prop:blockreduction} the compressed pair $X_-<X_+$, started at $a<b$, is independent of the compressed singleton $s+N$. For $(2,1)$ the pair is weaker, $(a,b)=(0,1)$ and $s=j$. For $(1,2)$ it is stronger, $s=0$ and $a,b$ are the two sites other than $j$. Let the singleton be at compressed site $x$. In the first case a pair member at compressed site $u$ has physical site $u+\one\{u\ge x\}$; in the second the singleton has physical site $x+\one\{X_-\le x\}+\one\{X_+\le x+1\}$. Hence $\mathsf L_t=\one\{X_-\le x_1\}+\one\{X_+\le x_2\}$, where $(x_1,x_2)=(x-1,x-1)$ for $(2,1)$ and $(x_1,x_2)=(x,x+1)$ for $(1,2)$. With $F_\pm$ the distribution functions of $X_\pm$ from Lemma~\ref{lem:pair},
\[
 K_t(j,\cdot)=\E\,g_t(s+N),\qquad
 g_t(x)=\bigl(1-F_-(x_1),\ F_-(x_1)-F_+(x_2),\ F_+(x_2)\bigr).
\]
Through $F_\pm$ the vector $g_t$ also depends on the starting pair $(a,b)$, which we do not display.

\emph{Normal approximation.} Let $G$ be standard normal with density $\phi$, distribution function $\Phi$ and upper tail $\bar\Phi=1-\Phi$. Put $\widehat N=(N-t)/\sqrt t$ and $z=(x-t)/\sqrt t$. For fixed $c$ and uniformly in $x\in\Z$,
\begin{equation}\label{eq:normalapprox}
\begin{gathered}
 F_c(x)=\Phi(z)+O(t^{-1/2}),\qquad
 \sqrt t\,f_c(x)=\phi(z)+O(t^{-1/2}),\\
 t^{-1/2}\,\E[(c+N-x)_+]=\E[(G-z)_+]+O(t^{-1/2}),
\end{gathered}
\end{equation}
and the last estimate also holds for $(x-c-N)_+$ and $(z-G)_+$. The first is the Berry--Esseen bound for the sum of independent Poisson variables used in the proof of Theorem~\ref{thm:main}. The third follows from the Wasserstein bound in the proof of Theorem~\ref{thm:main}, since positive parts are $1$-Lipschitz. For the second, Fourier inversion compares the exponents $t(e^{iu/\sqrt t}-1-iu/\sqrt t)$ and $-u^2/2$ on $|u|\le\pi\sqrt t$; they differ by at most $|u|^3/(6\sqrt t)$ and both have real part at most $-2u^2/\pi^2$. This is the case $c=0$; for fixed $c$, apply the three estimates at $x-c$ and use the bounded derivatives of $\Phi$ and $\phi$. Since $\E[(G-z)_+]=\phi(z)-z\bar\Phi(z)$ and $\E[(z-G)_+]=\phi(z)+z\Phi(z)$ are $O(1+|z|)$, inserting \eqref{eq:normalapprox} in \eqref{eq:paircdfs} gives
\begin{equation}\label{eq:rowapprox}
 g_t(x)=g(z)+O\bigl((1+|z|)/\sqrt t\bigr).
\end{equation}
For blocking, $g=(1-\Phi_{\rm bl},\,\Phi_{\rm bl}-\Phi,\,\Phi)$ with
\[
 \Phi_{\rm bl}(z)=1-\bar\Phi(z)^2+\phi(z)[\phi(z)-z\bar\Phi(z)],\qquad
 \Phi_{\rm bl}'(z)=\phi(z)[(1+z^2)\bar\Phi(z)-z\phi(z)].
\]
For pushing, $g(z)$ is the blocking vector at $-z$ in reverse order. In both cases $g$ and $z\mapsto zg(z)$ are Lipschitz.

\emph{Limit and unit shift.} By \eqref{eq:rowapprox} and the Wasserstein bound,
$K_t(j,\cdot)=\mu_\infty+O(t^{-1/2})$ with $\mu_\infty=\E g(G)$. For a fixed starting pair, the Poisson identity $f_0(k-1)=kf_0(k)/t$ gives
\begin{equation}\label{eq:singletonslope}
 \E g_t(s+1+N)-\E g_t(s+N)
 =\frac{\E[\widehat N\,g_t(s+N)]}{\sqrt t}
 =\frac{v}{\sqrt t}+O(t^{-1}),\qquad v=\E[G\,g(G)],
\end{equation}
where the last step uses \eqref{eq:rowapprox}, $\E\widehat N^2=1$ and the Wasserstein bound once more. For blocking, $\E\Phi(G)=\tfrac12$, $\E\bar\Phi(G)^2=\tfrac13$, $\int_\R\phi^3=1/(2\pi\sqrt3)$ and $\int_\R z\phi^2\bar\Phi=-\tfrac12\int_\R\phi^3$ give $\E[1-\Phi_{\rm bl}(G)]=\tfrac13-\tfrac32\int_\R\phi^3=p_{\rm L}$, so $\mu_\infty=(p_{\rm L},p_{\rm M},\tfrac12)$. Gaussian integration by parts gives
\[
 \E[G\Phi(G)]=\int_\R\phi^2=\frac1{2\sqrt\pi},\qquad
 \E[G\Phi_{\rm bl}(G)]=\int_\R\phi\,\Phi_{\rm bl}'
 =\tfrac12\int_\R(1+z^2)\phi(z)^2\,dz=\frac3{8\sqrt\pi},
\]
using $\bar\Phi(z)+\bar\Phi(-z)=1$ and the oddness of $z\phi^3$. Hence $v=(-3,-1,4)/(8\sqrt\pi)$. For pushing, the symmetry $z\mapsto-z$ reverses $\mu_\infty$ and gives $v=(-4,1,3)/(8\sqrt\pi)$.

\emph{Dependence on the initial word.} For $(2,1)$ the pair starts at $(0,1)$ for every $j$ and $s=j$, so \eqref{eq:singletonslope} gives $K_t(j,\cdot)=K_t(0,\cdot)+\ell_jv/\sqrt t+O(t^{-1})$ with $\ell=(0,1,2)$. For $(1,2)$ the pair starts at $(1,2),(0,2),(0,1)$ for $j=0,1,2$. Under blocking $F_+=F_b$ does not involve $a$, and under pushing $F_-=F_a$ does not involve $b$. By \eqref{eq:paircdfs}, moving $a$ to $a+1$ under blocking, or $b$ to $b+1$ under pushing, changes the other marginal at $x$ by
\[
 -f_a(x)[1-F_b(x)]+f_b(x+1)[1-F_a(x-1)]
 \qquad\text{or}\qquad
 -F_a(x)f_b(x)+f_a(x)F_b(x),
\]
respectively. Both are $O(t^{-1})$, since by \eqref{eq:normalapprox} the two products have the same leading term, $\phi(z)\bar\Phi(z)/\sqrt t$ in the first expression and $\phi(z)\Phi(z)/\sqrt t$ in the second. Up to $O(t^{-1})$ the row therefore depends on the starting pair only through $b$ under blocking and only through $a$ under pushing. Finally, the pair $(0,1)$ is the pair $(1,2)$ moved one site to the left, which has the same effect on the word as moving the singleton one site to the right. The three rows are thus again $K_t(0,\cdot)+\ell_jv/\sqrt t+O(t^{-1})$, with $\ell=(0,0,1)$ under blocking and $\ell=(0,1,1)$ under pushing.

\emph{Information.} Let $\bar\ell$ be the mean of $\ell_0,\ell_1,\ell_2$. The output law is $\mu_t=\mu_\infty+O(t^{-1/2})$, with positive limiting entries, and $K_t(j,\cdot)-\mu_t=(\ell_j-\bar\ell)\,v/\sqrt t+O(t^{-1})$. The second-order expansion of relative entropy used for Theorem~\ref{thm:main} gives
\[
 t I_{\boldsymbol n}^\alpha(t)
   =\frac12\,\Var(\ell_{\mathsf L_0})
      \sum_{k=0}^2\frac{v_k^2}{\mu_\infty(k)}+O(t^{-1/2}).
\]
The sum equals $\mathcal J$ under either rule, and the variance is $2/3$ for $(2,1)$ and $2/9$ for $(1,2)$.
\end{proof}

\section{Summary and open problems}\label{sec:discussion}

For distinct species, the compressed coordinates evolve as independent
rate-one Poisson walks given the initial configuration. If the law $\lambda$
of the initial configuration has finite support, then
$I(U;W_t)=C_U/t+O_{n,\lambda}(t^{-3/2})$ for every function $U$ of that
configuration (Theorem~\ref{thm:main}). For two particles of distinct species
at fixed sites with a uniform initial word, the ordering information decays as
$t^{-1}$ under long-range swaps and as $t^{-1/2}$ under ordinary multispecies
TASEP (Theorem~\ref{thm:two}). For consecutive initial sites and a uniform
initial word, the coefficient of the short-time $t\log(1/t)$ loss is symmetric
in the species multiplicities under the drop--push-type rule and can depend on
their order from weak to strong under the TASEP-type rule
(Theorem~\ref{thm:composition} and Corollary~\ref{cor:diversity}). For three
particles, the long-time coefficient for $(2,1)$ is three times that for
$(1,2)$ under either rule (Theorem~\ref{thm:repeated}).

\paragraph{Monotonicity.}
For $n\ge3$ distinct species, is $I(W_0;W_t)$ nonincreasing in $t$ for every
choice of fixed initial sites and every prior on $W_0$?
Proposition~\ref{prop:tv2} proves this for $n=2$ by showing that the later
channel factors through the earlier one. For general $n$, data processing
gives monotonicity for the full configuration, but the word alone hides the
gaps that affect later changes of order. For three distinct species we tested
$5{,}616$ combinations of initial sites and priors at $180$ times from
$10^{-3}$ to $10^{2.5}$; the largest computed increase between successive
times was below $10^{-15}$ nats in binary64 arithmetic (Supplement~S1).
For other inputs the information can increase: in the two-particle example
preceding Proposition~\ref{prop:tv2}, where the initial word is fixed and $U$
is a function of the initial gap, $I(U;W_0)=0$ while $I(U;W_t)>0$ for all
sufficiently large $t$.

\paragraph{Repeated species.}
Let the initial sites be consecutive and the initial word uniform, and let
$\boldsymbol n$ have at least two species, at least one of them repeated.
Does $tI_{\boldsymbol n}^{\alpha}(t)$ converge to a positive finite limit as
$t\to\infty$ for each $\alpha\in\{0,1\}$, and how does the limit depend on the
multiplicities, their priority order and the rule? Does the inequality
$I_{\boldsymbol n}^{1}(t)<I_{\boldsymbol n}^{0}(t)$, which
Corollary~\ref{cor:diversity} gives for small $t$, hold for every $t>0$?
For $(2,1)$ and $(1,2)$ it holds at all thirteen times plotted in
Figure~\ref{fig:late}(b). By Theorem~\ref{thm:repeated}, however, the two
rules have the same $t^{-1}$ coefficient, so the sign of the difference at
large times is decided by terms beyond leading order. A nonuniform prior can
reverse the inequality: for $\boldsymbol n=(1,2)$ and
$\rho_j=\PP(\mathsf L_0=j)$, the row expansions in the proof of
Theorem~\ref{thm:repeated} give
$\lim_{t\to\infty}tI(W_0;W_t)=\tfrac12\mathcal J\rho_2(1-\rho_2)$ under
blocking and $\tfrac12\mathcal J\rho_0(1-\rho_0)$ under pushing, and the
second limit is larger whenever $\rho_1>0$ and $\rho_0>\rho_2$.

\section*{Supplementary Materials}
Supplement~S1 (\path{ordering_information_supplement.zip}) contains the Python programs that
reproduce the numerical values, tables and figure data of this paper, with instructions for running them.

\section*{Author Contributions}
E.L. is the sole author and is responsible for the conception, mathematical arguments, numerical analysis and writing.

\section*{Funding}
{This research was funded by Nazarbayev University under Faculty-development competitive research grants program for 2024-2026 Grant 201223FD8822, EL.}

\section*{Data Availability Statement}
No external data were used. The programs in Supplement~S1 regenerate all numerical values,
tables and figure data reported in this paper.

\section*{Acknowledgments}
During the preparation of this manuscript, the author used ChatGPT Astra (OpenAI) and Claude Fable 5.2 (Anthropic) to verify proofs, search the literature, develop and verify Python code, produce figures and tables, and review and edit the text. The author has reviewed and edited all output and takes full responsibility for the mathematical arguments, computations, references, and final text.

\section*{Conflicts of Interest}
{The author declares no conflicts of interest.}

\renewcommand{\bibfont}{\small}
\setlength{\bibsep}{4pt plus 1pt minus 1pt}
\bibliographystyle{unsrtnat}
\bibliography{refs}

\end{document}